\documentclass[twocolumn]{aastex63}

\usepackage[utf8]{inputenc}
\usepackage{natbib}
\usepackage{graphicx}

\usepackage{textcomp}
\usepackage{gensymb}
\usepackage{enumitem}
\usepackage{xcolor}
\usepackage{hyperref}
\usepackage{lineno}
\begin{document}
\graphicspath{{./}{figures/}}
%\linenumbers %need to un-comment this line before submission

%\submitjournal{The Astronomical Journal}
\received{27 October 2021}
\revised{14 February 2022 and 16 March 2022}
\accepted{17 March 2022 to the Astronomical Journal}

\newcommand{\lah}[1]{\textcolor{brown}{#1}}
\newcommand{\lahcomm}[1]{\textcolor{brown}{COMMENT: #1}}

\title{A ZTF Look at Optical Variability of Young Stellar Objects \\ in the North America and Pelican Nebulae Complex}

% This title is a nod to one of my favorite papers: "A NICER look at the state transitions of the black hole candidate MAXI J1535-571 during its reflares"%

\correspondingauthor{Lynne A. Hillenbrand} 
\email{lah@astro.caltech.edu} 

\author{Lynne A. Hillenbrand}
\affiliation{Department of Astronomy, California Institute of Technology, Pasadena, CA 91125, USA}

\author[0000-0002-2363-2487]{Thaddaeus J. Kiker}
\affiliation{Harriton High School, Bryn Mawr, PA 19010, USA}
\affiliation{Sunny Hills High School, Fullerton, CA 92833, USA}

\author{Miles Gee}
\affiliation{Harriton High School, Bryn Mawr, PA 19010, USA}

\author{Owen Lester}
\affiliation{Harriton High School, Bryn Mawr, PA 19010, USA}

\author{Noah L. Braunfeld}
\affiliation{Harriton High School, Bryn Mawr, PA 19010, USA}

\author[0000-0001-6381-515X]{Luisa M. Rebull}
\affiliation{Infrared Science Archive (IRSA), IPAC, 1200 E. California Blvd., California Institute of Technology, Pasadena, CA 91125, USA}

\author[0000-0002-0631-7514]{Michael A. Kuhn}
\affiliation{Department of Astronomy, California Institute of Technology, Pasadena, CA 91125, USA}

%%%%%%%ABSTRACT%%%%%%%%
\begin{abstract}
We present a study of 323 photometrically variable young stellar objects that are likely members of the North America and Pelican (NAP) nebulae star forming region. 
%Our work constitutes the largest set of YSOs studied with these metrics to date. 
To do so, we utilize over two years of data in the $g$ and $r$ photometric bands from the Zwicky Transient Facility (ZTF). We first investigate periodic variability, finding 46 objects ($\sim$15\% of the sample) with significant periods that phase well, and can be attributed to stellar rotation. We then use the quasi-periodicity (Q) and flux asymmetry (M) variability metrics to assign morphological classifications to the remaining aperiodic light curves.  Another $\sim$39\% of the variable star sample beyond the periodic sources are also flux-symmetric, but with a quasi-periodic (moderate $Q$) or stochastic high $Q$) nature.   Concerning flux-asymmetric sources, our analysis reveals  $\sim$14\% bursters (high negative $M$) and $\sim$29\% dippers (high positive $M$).  We also investigate the relationship between variability slopes in the $g$ vs $g-r$ color-magnitude diagram, and the light curve morphological classes. Burster-type objects have shallow slopes, while dipper-type variables tend to have higher slopes that are consistent with extinction driven variability.  Our work is one of the earliest applications of the $Q$ and $M$ metrics to ground-based data. We therefore contrast the $Q$ values of high-cadence and high-precision space-based data, for which these metrics were designed, with $Q$ determinations resulting from degraded space-based lightcurves that have the cadence and photometric precision characteristic of ground-based data. 
\end{abstract}

\keywords{Young star clusters(1833);
Pre-main-sequence stars (1290);
T Tauri stars (1681); 
Irregular variable stars(865); 
Stellar rotation(1629);
Time domain astronomy(2109)} 

%%%%%%%INTRODUCTION%%%%%%
\section{Introduction} \label{sec:intro}

The Zwicky Transient Facility \citep[ZTF;][]{Kulkarni2018} has been used prolifically in contemporary transient detection and characterization efforts. Astrophysical transients observed and/or discovered by ZTF include: tidal disruption events \citep[e.g.][]{vanVelzen2021}, candidate electromagnetic counterparts to gravitational wave events \citep[e.g.][]{Graham2020}, type Ia supernovae \citep[e.g.][]{Yao2019,Miller2020,Bulla2020}, dwarf novae \citep[e.g.][]{Soraisam2021}, cataclysmic variables \citep[e.g.][]{Szkody2020}, AM CvN systems \citep{vanRoestel2021b}, general stellar variables \citep[e.g.][]{vanRoestel2021a,Roulston2021}, active galactic nuclei \citep[e.g.][]{Frederick2019,Ward2021}, and solar system bodies such as comets and asteroids \citep[e.g.][]{Ye2020a,Ye2020b}. 

While photometric variability has long been known as a ubiquitous characteristic of young stellar objects \citep{Joy1945}, no study to date has utilized ZTF for investigating the general variability charactistics of young stars based on a significantly large sample. ZTF data have, however, contributed to investigations of young stellar objects (YSOs) in several studies focused on individual sources \citep[e.g.][]{Dahm2020, Hodapp2020, Hillenbrand2022}.  The plethora of multi-year, nearly nightly cadence, and multi-color ZTF observations is ideally suited for studying the variability of optically visible YSOs. 

\begin{figure*}[ht]
\includegraphics[width=0.95\textwidth]{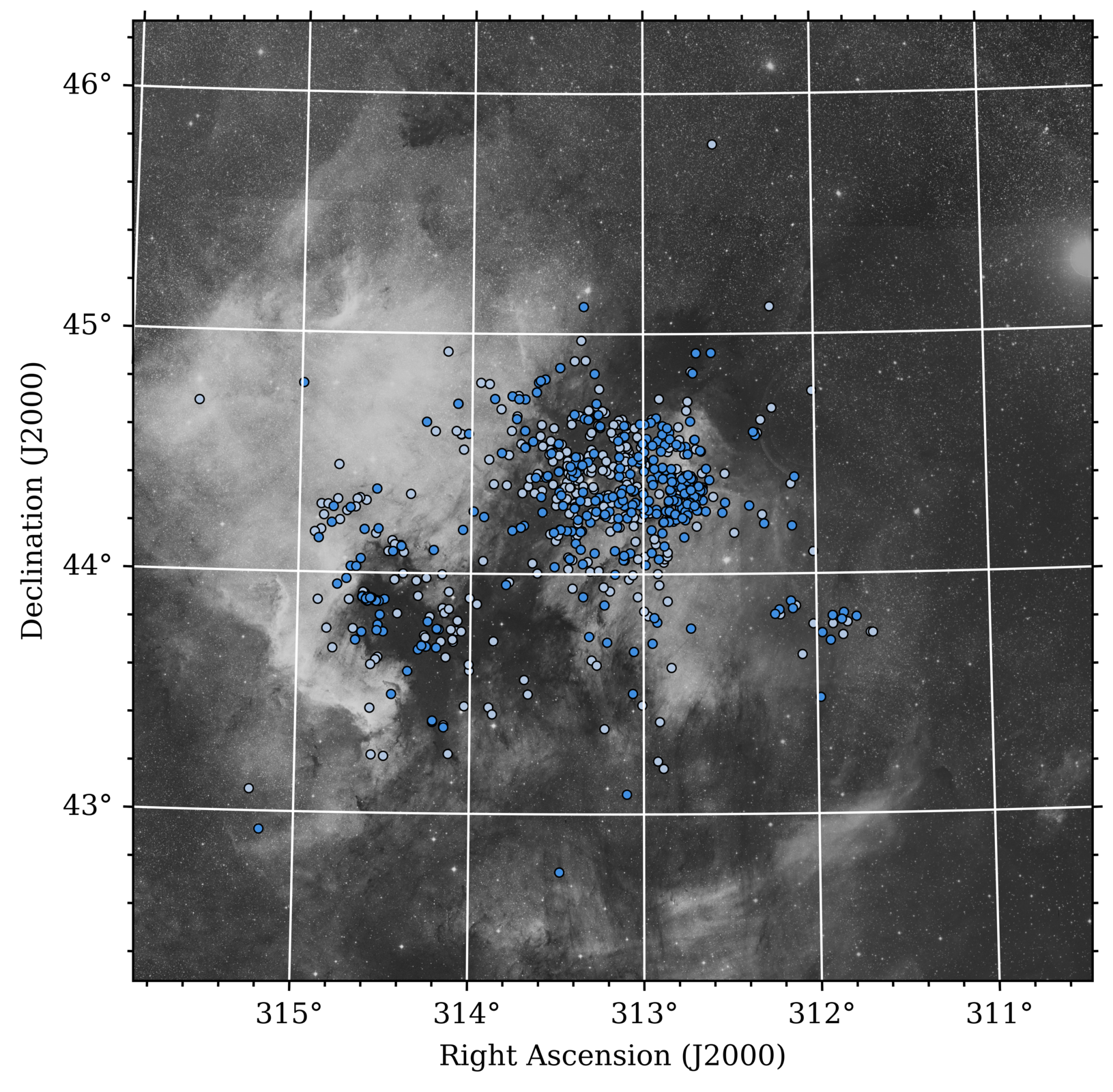}.
\centering
\caption{Full sample of selected likely members of the North America (left) and Pelican (center) nebulae star-forming complex plotted over a DSS2-R red photographic image of the region. %% CHeck if this is the right one? might be 535, not sure
Our final sample of 323 variable sources from ZTF is in darker blue, with the remaining likely member sources that do not meet all of the applied ZTF photometric cuts represented in lighter blue.}
\label{fig:skyview}
\end{figure*}

Our study targets a sample of YSOs in the vicinity of the North America and Pelican (hereafter NAP) nebulae \citep{Reipurth2008}.
Initial interest in this region as a laboratory for studying the evolution and variability of YSOs was piqued after \cite{Herbig1958} identified and characterized numerous YSOs based on H$\alpha$ emission lines. More recently, \cite{Guieu2009} and \cite{Rebull2011} identified more than two thousand candidate YSOs in the region based on infrared excess emission using data from Spitzer \citep{Werner2004} and 2MASS \citep{Skrutskie2006}. Focussing on the L935 dark cloud region,  \cite{Armond2011} identified new H$\alpha$ emitters and Herbig-Haro flows, and \cite{Damiani2017} studied x-ray emitters. These catalogs have been of interest in recent years for other studies investigating various aspects of the young stars in the region, including kinematics \citep{Kuhn2020} and HR diagrams \citep{Fang2020}. 

Considering previous variability studies, \cite{Findeisen2013} identified 43 large-amplitude aperiodic variables with distinct bursting or fading behavior, which is believed to be driven by variable accretion and extinction processes, respectively \citep[e.g.][]{Herbst2002, Cody2018}. The results of long-term photometric monitoring observations of the region have been reported by \cite{Poljancic2014}, \cite{Ibryamov2015}, and \cite{Ibryamov2018}, providing detailed analyses of small samples of stars based on decades of multi-color data from numerous observatories.  \cite{Bhardwaj2019} found 56 periodic variables in the Pelican Nebula region, visually classifying an additional 11 variables as non-periodic while \cite{Froebrich2021} identified 59 periodic variables, many of which overlap the \cite{Bhardwaj2019} sample.

In this paper, we utilize the flux asymmetry and quasi-periodicity metrics of \cite{Cody2014} to quantitatively describe variability classes of YSOs in the region, greatly expanding upon the results of previous variability studies. In doing so, we present the largest single sample of YSOs classified by the flux asymmetry and quasi-periodicity metrics to date. 
% yes, true, original cody was 162 sources in 2264
We connect the light curve classifications with analysis of color-magnitude variability trends, which helps to provide a more comprehensive understanding of the nature of the variability. 

The remainder of this paper is structured as follows. We describe our stellar sample in Section \ref{sec:sample} and the observations used as well as data selection and handling in Section \ref{sec:observations}. The identification of variability and the search for periodicities are discussed in Section \ref{sec:var}. In Section \ref{sec:varclass} we present our light curve classification and color-magnitude results. In Section \ref{sec:disks} we consider relationships of the variability patterns to the presence of circumstellar disks.  We discuss our results in the context of other recent studies in Section \ref{sec:discussion}, and summarize our findings in Section \ref{sec:conclusion}. Several appendices present: the full light curve set, the results of testing performance of one of the variability metrics (quasi-periodicity) with data cadence and precision,
and our findings on the periodograms for two particularly interesting sources.  

%%%%SAMPLE SELECTION%%%%%%%
\section{Sample Selection} \label{sec:sample}
Our sample consists of YSOs in the North America and Pelican Nebulae that have been promoted as members based on kinematic, spectroscopic, and H-R diagram selection criteria. 

The starting list for this investigation was created by cross-matching all 580 objects from Table 4 in \cite{Fang2020} with all 395 objects from Table 2 in \cite{Kuhn2020} using a match radius of $0.5\arcsec$.  The combined sample of 696 sources includes every source in either of those tables, 
with 301 uniquely from \cite{Fang2020} and 116 uniquely from \cite{Kuhn2020}. 

In \cite{Kuhn2020}, membership was assessed using Gaia kinematics for all candidate members of the North America and Pelican Nebula complex previously suggested as such in the literature.  
In \cite{Fang2020}, membership for stars with optical spectroscopic data was assessed from a combination of indicators. These included stellar activity manifest through X-rays and accretion-related optical emission lines, infrared excess indicative of circumstellar dust, lithium absorption, and parallax and proper motion criteria.

Figure~\ref{fig:skyview} shows the spatial distribution of the initial sample as well as the reduced, final sample discussed below.

%%%%%%%%%%OBSERVATIONS%%%%%
\section{ZTF Photometry} \label{sec:observations}
ZTF \citep{Bellm2019} employs the 48-inch Schmidt telescope at Palomar Observatory and possesses a camera with a 47 deg$^2$ field of view. It efficiently surveys the northern sky, cataloging more than a billion objects since first light in 2018 \citep{Masci2019}. ZTF serves as the current standard for rapid cadence, ground based photometric surveys as well as a test bed for techniques to be used in upcoming next generation surveys such as LSST \citep{Graham2019}. 

In this paper, we analyze data from the fourth public ZTF data release (ZTF DR4), which corresponds to more than two years of data taken in the NAP field between 28 March 2018 and 6 June 2020 ($58205 \leq$ MJD $\leq 59028$).  We focus our analysis on the ZTF $g$-band ($\lambda_{\rm eff}=4722$ \AA) and $r$-band ($\lambda_{\rm eff}=6340$\AA) data, as the long-term $i$-band survey mainly covers high galactic latitude fields. We refer the reader to \cite{Bellm2019} and \cite{Masci2019} for in depth descriptions of the ZTF observing and image processing systems.  The photometry is calibrated to the PanSTARRS photometric system (with 
$g$ having $\lambda_{\rm eff}=4811$ \AA\ and $r$ having $\lambda_{\rm eff}=6156$\AA) and reported in AB magnitudes. 

Two objects were removed from the sample 
(2MASS J20505838+4414444 near FHK 47 = V1597 Cyg A,
and Gaia DR2 2163143481019730688 near FHK 67) 
because they are the fainter members of close pairs, and the photometry is spatially unresolved in ZTF.
We additionally note that while 2MASS J20593270+4408353 %Gaia DR2 2162231676644797824 
and FHK 485 are spatially resolved in ZTF images, their photometry and hence lightcurves appear cross-contaminated.

The median cadence of the ZTF light curves in the NAP region is $\sim1$ day. 
The observations were generally taken nightly, apart from seasonal gaps of $\sim30$ days plus additional short gaps due to adverse weather. 
We ignore observations affected by clouds and/or the moon (these observations have catflag = 32768). 
We also ignore the observations between 58448 and 58456 MJD as these epochs encompass the ZTF high-cadence experiments \citep{Kupfer2021}, which have some noticeably poor photometry in the NAP region due to bad weather, which adversely impacts period searches and color-magnitude analysis. 

\begin{table}
    \centering
    \caption{Remaining sample size following application of selection criteria.}
    \begin{tabular}{lr}
    \tableline
    \tableline
    Criterion & \# in Sample \\
    \tableline
    Initial sample of likely NAP members & 696 \\ 
    Spatially resolved in ZTF & 694\\ 
    Brightness $g<20.8$ or $r<20.6$ mag & 410 \\
    Epochs $>30$ for both $g$ and $r$ & 392 \\
    Variability level $\nu > 15$th percentile & 323\\
    \tableline
    \end{tabular}
    \label{tab:cuts}
\end{table}

\begin{figure}[!htbp]
\includegraphics[width=0.5\textwidth]{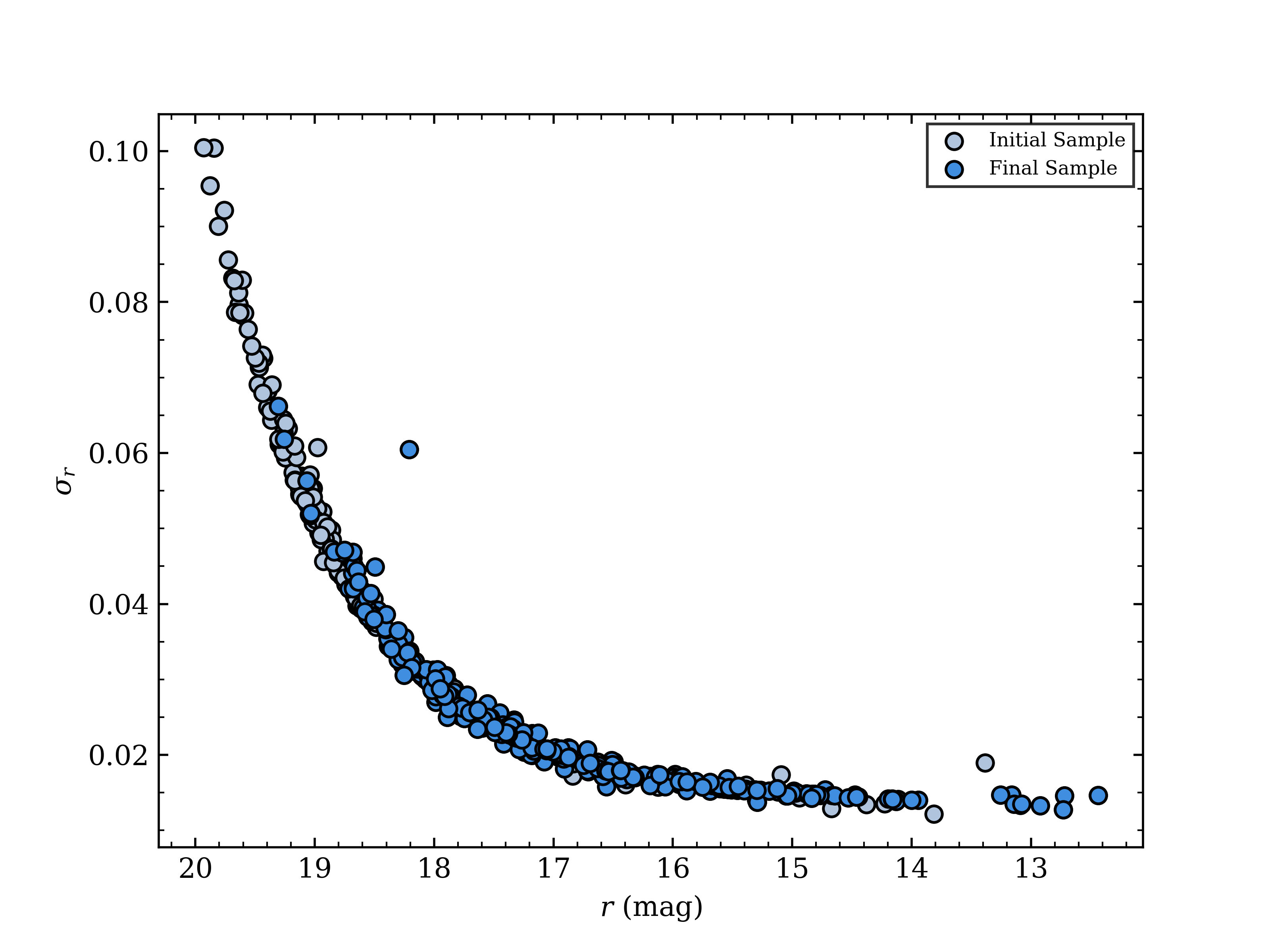}
\centering 
\caption{Mean $1\sigma$ photometric uncertainty as a function of mean $r$ magnitude, calculated from the ZTF DR4 data products. 
Objects in our final variable star sample are shown in darker blue, and those not meeting all photometric cuts in lighter blue.
The outlier at $r\approx18$ and $\sigma_r\approx0.06$ is FHK 96,  %$\mathrm{GDR1}\mathrm{\;} 2162947420052773760$, 
a source that undergoes an abrupt dimming event of $>4.5$ mag approximately halfway through the observing window.} 
\label{fig:magmagerr}
\includegraphics[width=0.5\textwidth]{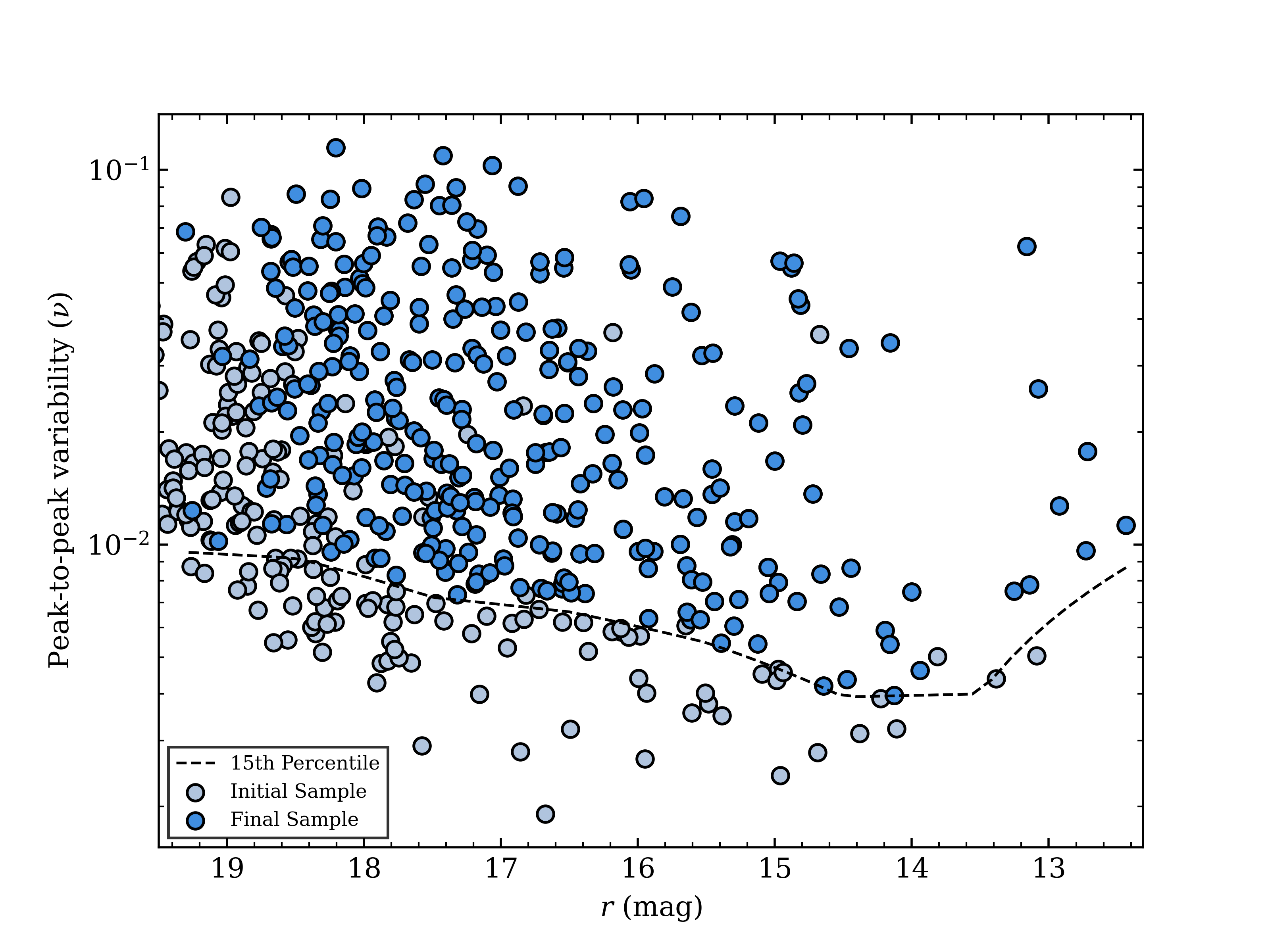}
\centering
\caption{Normalized peak-to-peak variability metric $\nu$ as a function of mean $r$ magnitude for our sample of 392 NAP stars that meet the cuts on magnitude and minimum number of observations. Objects in darker blue above the dashed $15$th percentile line are the 323 sources included in our variable star sample, while those not included are in lighter blue.}
\label{fig:magvsnu}
\end{figure}

For our analysis, we restrict the sample to objects with mean $g < 20.8$ mag or mean $r < 20.6$ mag over the entire time series. We further require at least 30 observations in both the $g$-band and $r$-band (after the application of a 5$\sigma$ clip of potential outliers from the median magnitude of the light curve). A summary of the cuts applied to our initial sample and its reduction to our final sample for light curve analysis is provided in Table \ref{tab:cuts}. Figure \ref{fig:magmagerr} shows the photometric precision of the $r$-band observations as a function of median $r$ magnitude for objects in our final sample.

\begin{table*}[ht] 
    \caption{Properties of variable stars in our NAP sample. 
    A full, machine-readable, version of this table is available electronically.}
    %\footnote{\url{https://github.com/HarritonResearchLab/NAPYSOs/tree/main/results}}}
    \centering
    \begin{tabular}{ c r r r r r r r r c c r }
    \tableline
    \tableline
    Identifier & R.A. & Dec. & Timescale& Q & M & $\nu$ & $<r>$ & Primary & Secondary & CMD \\
         &    &     & or Period&   &   &       &     & Var. Class & Var. Class & Angle \\
     & (deg) & (deg) & (d) & & & & (mag) &  & & (deg) \\
    \tableline
    V752 Cyg &  314.63997 &  44.05985 &  6.652 &  0.93 &  -0.05 &  0.0824 &  16.06 &    S &    - &  67.6 \\
    V1716 Cyg &  314.52548 &  43.88363 &   4.140 &  0.77 &  -0.49 &  0.0179 &  16.54 &    B &    P &  63.2 \\
    V1701 Cyg &  312.75373 &  44.53047 &  6.937 &  0.92 &   0.81 &   0.0840 &  15.96 &  APD &    - &  80.5 \\
    V1597 Cyg  &  312.74346 &  44.24561 &  3.422 &  0.86 &  -0.12 &  0.0087 &  15.05 &  QPS &    - &  48.3 \\
    V1492 Cyg &  312.77375 &  44.27562 &  10.48 &  0.88 &   0.39 &  0.0549 &  14.88 &  QPD &    - &  74.9 \\
    V1490 Cyg &  312.72325 &  44.35025 &  31.17 &  0.67 &   0.71 &  0.0571 &  14.96 &  QPD &    - &  81.9 \\
    LkHA 191 &  314.77429 &  43.95086 &      - &   0.90 &  -0.89 &  0.0113 &  12.44 &    B &    - &   7.5 \\
    LkHA 189 &  314.60004 &   43.89850 &  2.449 &  0.73 &   0.34 &  0.0118 &  15.57 &  QPS &    L &  65.3 \\
    LkHA 188 &  314.59921 &  43.88648 &  1.011 &  0.78 &   0.84 &  0.0406 &  13.16 &  QPD &    - &  74.3 \\
    LkHA 187 &  314.58975 &  43.89581 &  2.873 &  0.82 &   0.03 &  0.0095 &  16.32 &  QPS &    - &  59.6 \\
    \tableline
    \end{tabular}
     %can use \tablenotemark{a} in table and then here \tablenotetext{a}{for notes associated with marker a}
     \tablecomments{
     Timescale or Period = Lomb-Scargle periodogram peak that is identified with a period when the Primary Var. Class column is ``P"; otherwise interpreted as a variability timescale and not a strict period.  \\
     $Q$ = lightcurve quasiperiodicity metric defined in Equation 2. \\
     $M$ = lightcurve asymmetry metric defined in Equation 3. \\
     $\nu$ = variability metric defined in Equation 1. \\
     $<r>$ = mean magnitude over ZTF time series. \\
     Primary Variability Class = dominant behavior in the lightcurve, classified as one of:  
              P = strictly periodic; MP = multi-periodic; QPS = quasiperiodic, symmetric; QPD = quasiperiodic, dipping;
              APD = aperiodic, dipping; S = stochastic or aperiodic, symmetric behavior. \\
     Secondary Variability Class = additional, subordinate behavior exhibited in the lightcurve, using the same classification scheme as above. \\
     CMD Angle = fitted slope to the $g$ vs. $g-r$ color-magnitude diagram, reported only when the error is $<10\degree$.
     }
    \label{tab:properties}
\end{table*}

%%%%VAR. CLASSIFICATION%%%%
\section{Photometry Analysis} \label{sec:var} 

The $r$-band light curves, $g$ vs $g-r$ color-magnitude trends, and periodogram results are illustrated for a set of
representative sources in Appendix A; the full sample is available in an online figure set.

The results of our ZTF light curve analysis as described below appear in Table \ref{tab:properties}.
Source identifiers were adopted in the order: LkH$\alpha$ numbers from \cite{Herbig1958}, V* identifiers
from the General Catalog of Variable Stars (\citealt{Samus2017}; see also \citealt{Samus2021} for a history), 
FHK identifiers from \cite{Fang2020}, then catalog identifiers from 2MASS \cite{Cutri2003} and finally Gaia DR2 \cite{GC2018}. 

Photometric variability amplitude, and timescale or period (as appropriate) are discussed in this section.
Light curve asymmetry and waveform repeatability are presented in the next section.

\subsection{Variability Search} \label{sec:nu}

%we did investigate things like light curve std, iqr, 10-90, and von Neumann, and I remember there was a very strong correlation between std and nu in particular (after which we just went with nu). That was back in January perhaps. 

We utilize the normalized peak-to-peak variability metric

\begin{equation}\label{eq:defnu}
    \nu = \frac{(m_i-\sigma_i)_{\mathrm{max}}-(m_i+\sigma_i)_{\mathrm{min}}}{(m_i-\sigma_i)_{\mathrm{max}}+(m_i+\sigma_i)_{\mathrm{min}}}, 
\end{equation}

\noindent featured in \cite{Sokolovsky2017} to quantify variability amplitude for objects in our YSO sample. In this formula, $m_i$ represents a magnitude measurement and $\sigma_i$ represents the corresponding measurement error, with the minimum and maximum determined from the full light curve.  We eliminated possible outliers with the application of a $5\sigma$ clip to all light curves, before calculating $\nu$; this is necessary for the metric to be considered a sensitive variability indicator. For each $r$-band light curve in our sample that meets the measures described in Section \ref{sec:observations}, we calculate $\nu$. 
We note that $\nu$ correlates well with standard deviation, with values that are about a factor of 6 smaller.

Our final sample for variability analysis includes objects with $\nu$ greater than the $15$th percentile of $\nu$ as a function of magnitude. 
Although not a rigorously justified cutoff, this criterion ensures that we are studying the fractionally larger amplitude variables at each brightness level.
Figure \ref{fig:magvsnu} depicts the selection of sources based on $\nu$, and Table \ref{tab:cuts} shows the effect of the $\nu$ cut on our final sample size.

\subsection{Period Search}  \label{sec:periodic}

We use the Astropy implementation of the Lomb-Scargle periodogram \citep{VanderPlas2012, VanderPlas2015} to search for periodic signals in the $\sim$825 day ZTF data stream. Possible periods between 0.5 and 250 days are considered. 

We compute the periodogram for every object in the variable sample, flagging periods between 0.5--0.51, 0.98--1.02, 1.96--2.04, and 26--30 days for further analysis to avoid the most common aliases associated with the solar and sidereal days, as well as the lunar cycle  \citep[following][]{Rodriguez2017,Ansdell2018}.  In particular, the window function shows that we have the potential for extremely strong signals at frequencies corresponding to 1.000 and 0.997 day periods. To account for additional potentially aliased periods, we flag periods with half or double multiples that fail at least one of the aforementioned alias checks.   We then compute the $90\%$, $95\%$, and $99\%$ confidence false alarm probability levels, and consider the period corresponding to the highest power peak greater than the $99\%$ confidence false alarm probability (FAP $<  0.01$) to be significant \citep[following][]{Messina2010}. 
While FAP values may not be as reliable as formal hypothesis tests, they are sufficient for our purpose here, which is to identify potential variability timescales for visual examination and further assessment.

We repeat the above procedure with $g$ band data for objects lacking a significant period in $r$. If a significant period is found in $g$, we return this period for the source. 

For each object, the full light curve was phased at the four most significant peaks, and visually examined to assess the dispersion across phase.  We also marked periods corresponding to beats of the $\sim1$ day sampling interval to assess the expected fourier patterns.   The periodograms are quite clean in terms of signal-to-noise. In the vast majority of cases, the highest periodogram peak was retained as correct, with recognizable harmonic aliasing and beats having lower power.  
In some cases, a longer period with a slightly less prominent peak was retained as the most likely true period based on better explanation of the beat patterns, and improved phase dispersion minimization.  
We call attention to two particularly interesting and non-intuitive examples in Appendix B.

Objects with significant Lomb-Scargle periodogram peaks in either $r$ or $g$ that passed our vetting procedures are reported in Table~\ref{tab:properties}. We have labelled the relevant column as containing both timescales and true periods. In cases where the value of $Q$ (quasi-periodicity metric; see next section) is high, the periodogram peaks are more appropriately thought of as timescales, whereas when $Q$ is low, the light curve is truly periodic in the sense of strictly repeating, with low residual dispersion in the phased light curve.  In this latter case, there is also a ``P" in the column containing the primary variability classification.  A few sources have multiple significant real (non-alias or beat) peaks in the periodogram, and are given ``MP" classifications.
%see Table~\ref{tab:mp_objs} for additional information.
%with further discussion of two oddly long-period, multi-periodic objects appearing in Section~\ref{slowRotators}. 

\begin{figure}[htp]
    \centering
    \includegraphics[scale=0.6]{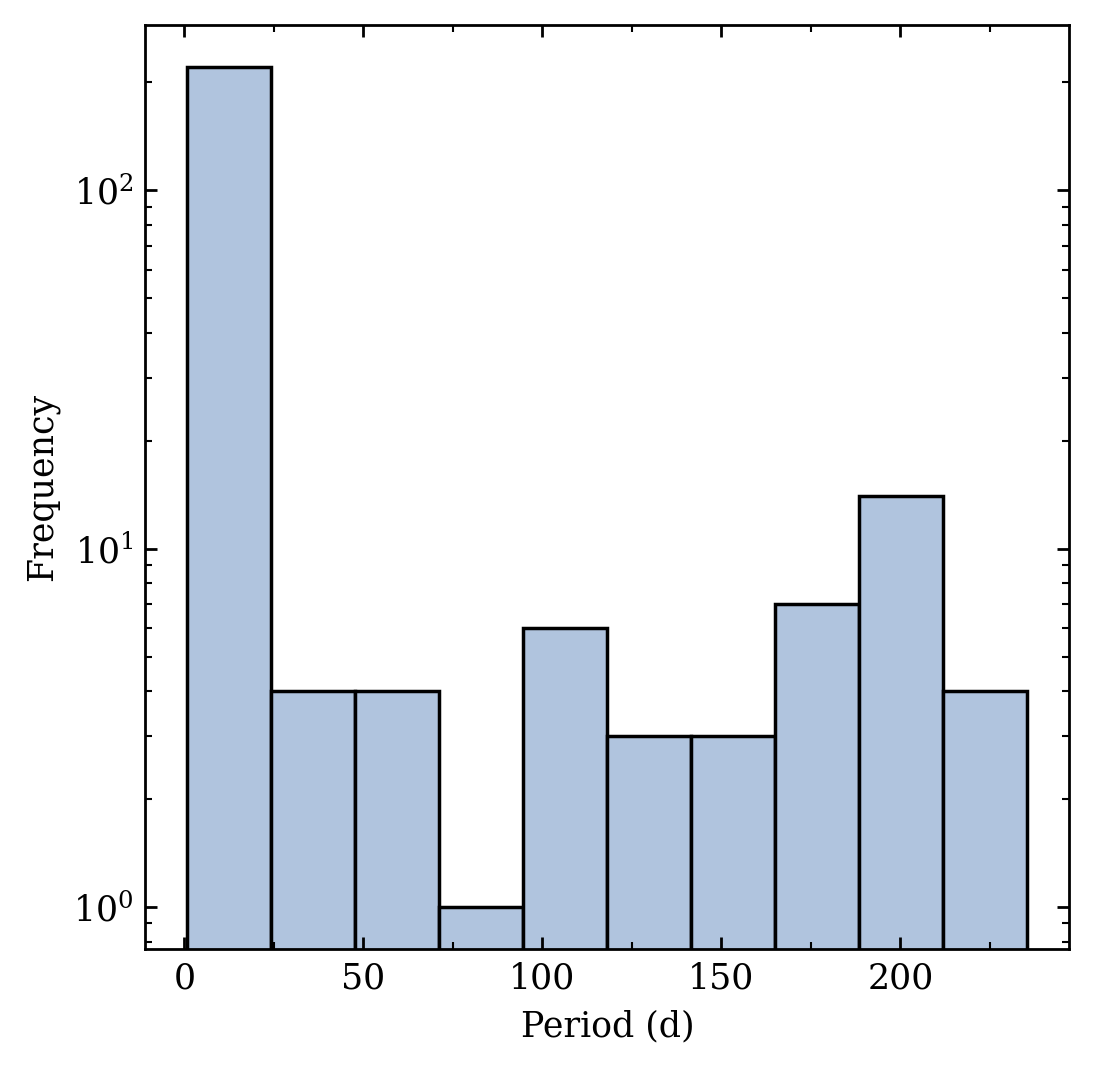}
     \includegraphics[scale=0.6]{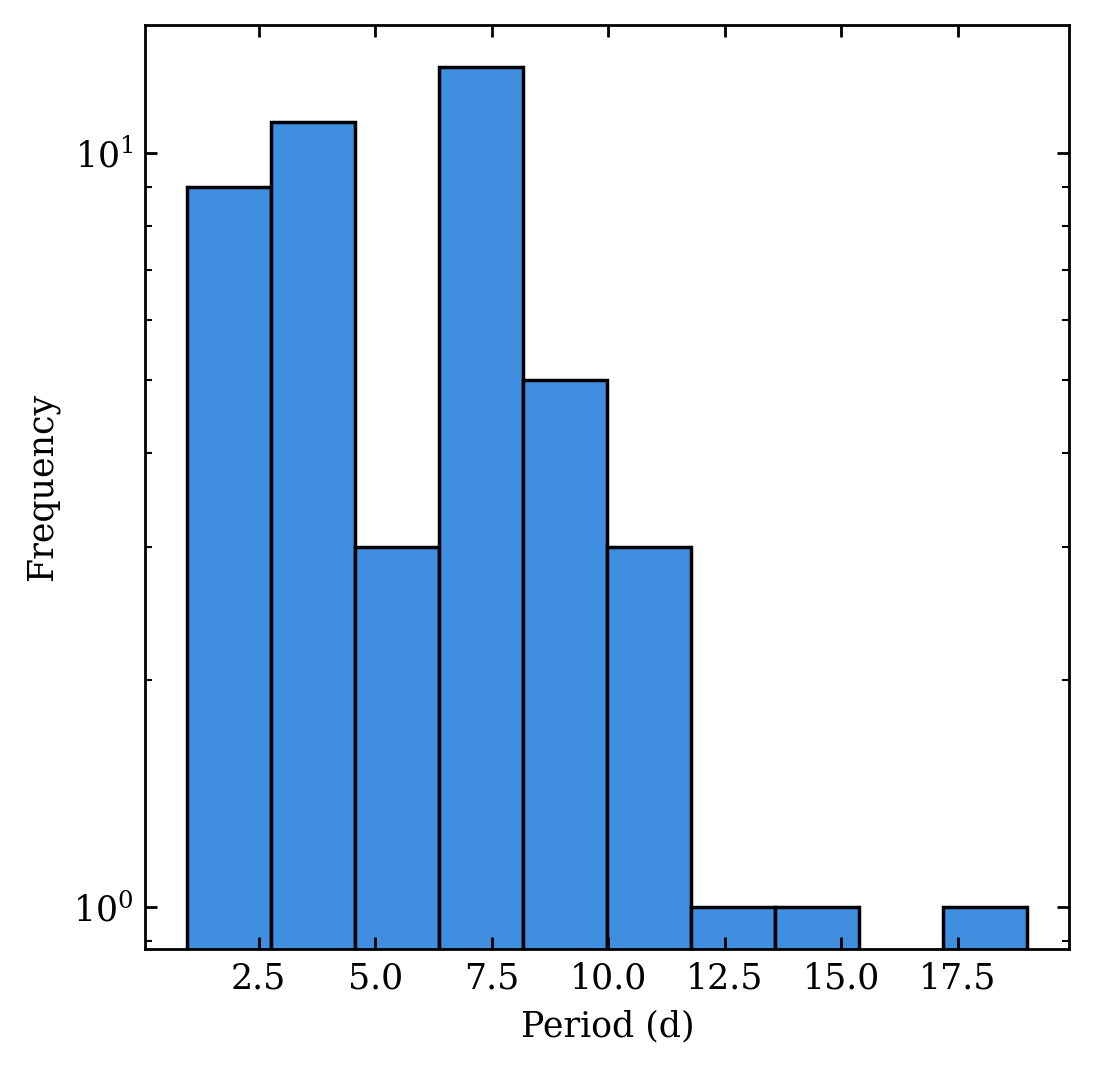}
    \caption{Distribution of significant Lomb-Scargle periodogram peaks.  Top panel shows shows all timescales from Table \ref{tab:properties}, regardless of light curve classification.
    Bottom panel shows only stars labelled as ``P" in Table \ref{tab:properties}, which are those that phase well to their periodogram peaks and likely have their origin in spot pattern modulation due to stellar rotation.  As expected, the true stellar rotation periods are concentrated
    towards the shorter time scales, with values of a few days to a few weeks.}
    \label{periodsdist}
\end{figure}

The distribution of periods for the sources identified as strictly periodic is shown in the lower panel of Figure~\ref{periodsdist}. 
There are 46 such stars, constituting $\sim$15\% of the variable sample with reported timescales or periods in Table~\ref{tab:properties}.  Their period distribution has a bimodal peak, as well as a skew towards longer periods.  Both features are consistent with the populations of fast and slow rotators discussed in previous young star literature, e.g. \cite{Gallet2013}.

Figure~\ref{periodsdist} also shows, in the upper panel, the full range of variability timescales in the sample. 
This distribution exhibits a strong peak on the same
few week timescales covered by the purely rotationally variable stars in the lower panel. 
The peak is significantly higher though, indicating a large quasi-periodic and aperiodic population 
with additional physical processes that are occurring on similar timescales. 
The timescale distribution is relatively flat out to longer times.  
The longer variability patterns, from weeks to months, can be seen in the light curves themselves.  
We emphasize that, even though these timescales were detected and measured using Lomb-Scargle periodogram techniques, 
they really are approximate timescales for photometric variability, and not true periods.  
Particularly in cases of high $Q$ values (see below), these timescales are less trustworthy.
Other methods for quantifying timescales in aperiodic light curves are discussed in \cite{Findeisen2015}. 

One particular source with a credible long period or timescale is V1490 Cyg, which is classified as a quasi-periodic dipper (QPD)  with a timescale of 31 days. In this case, the phased light curve shows substantially variable depth dips that are more consistent with material orbiting in a disk, than with e.g. a stellar rotation period.

We also performed a Lomb-Scargle period search on the $g-r$ color curves, with colors  created from the individual $g$ and $r$ time series as described below in \S5.2.  Few objects were found with significant periodicity in their colors.
Among them, FHK 216 and FHK 442 are examples with periodic color variations. While their color curves 
do not phase particularly well, the color periods are the same as the single-band periods, reported in Table~\ref{tab:properties}.
FHK 216 has a color-magnitude slope similar to the reddening vector while FHK 442 is slightly shallower.

%NOTES:  In YSOVAR it was common to find P(color) approximately equal to P(single band), 
%but also common to find a P(single band) and no P(color). 
%I don’t remember any P(color) without P(single band), but it has been a while.

%%RESULTS%%%
\section{Light Curve Classification} \label{sec:varclass}
\subsection{$Q$ and $M$ Variability Metrics}\label{5.1}

\begin{figure*}[ht]
\includegraphics[width=\textwidth]{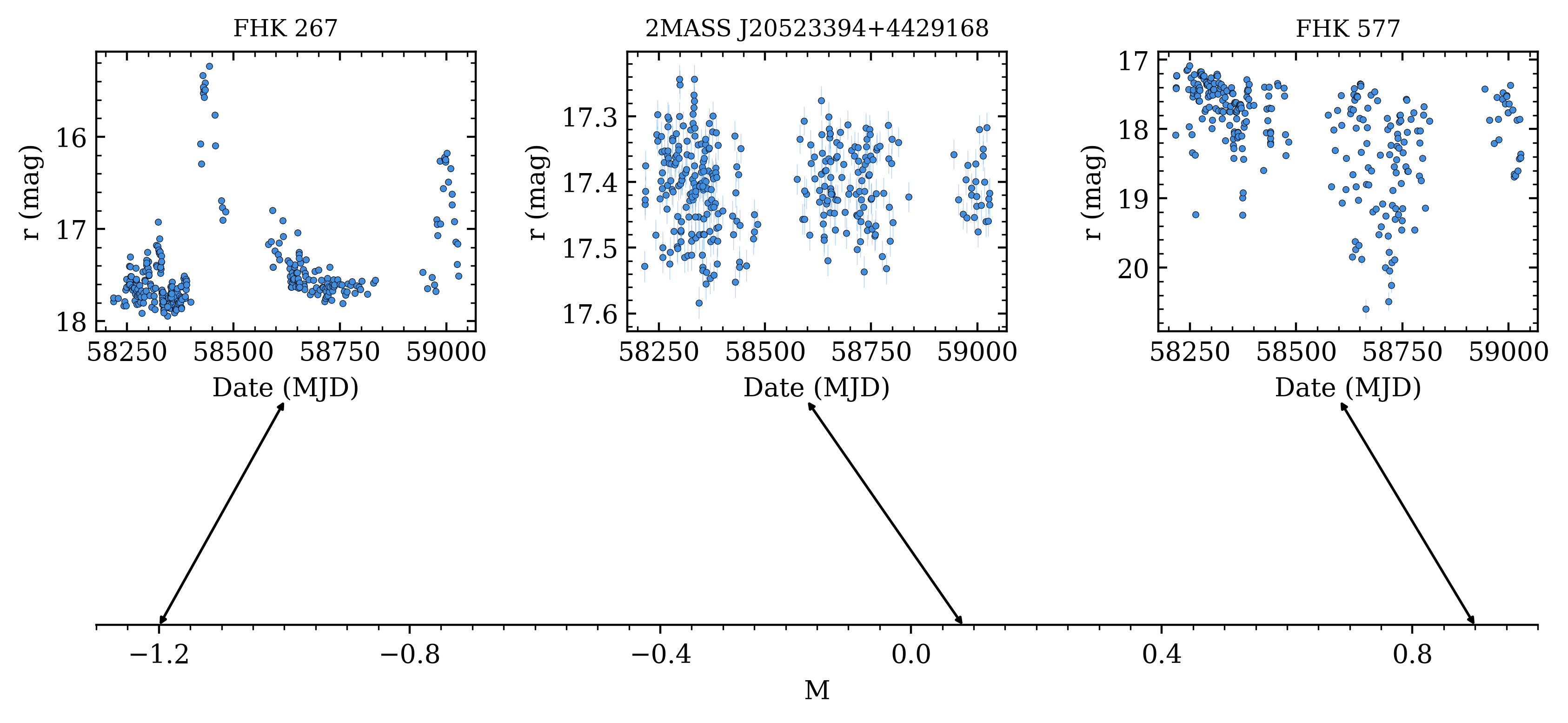}
\vskip 0.5truein
\includegraphics[width=\textwidth]{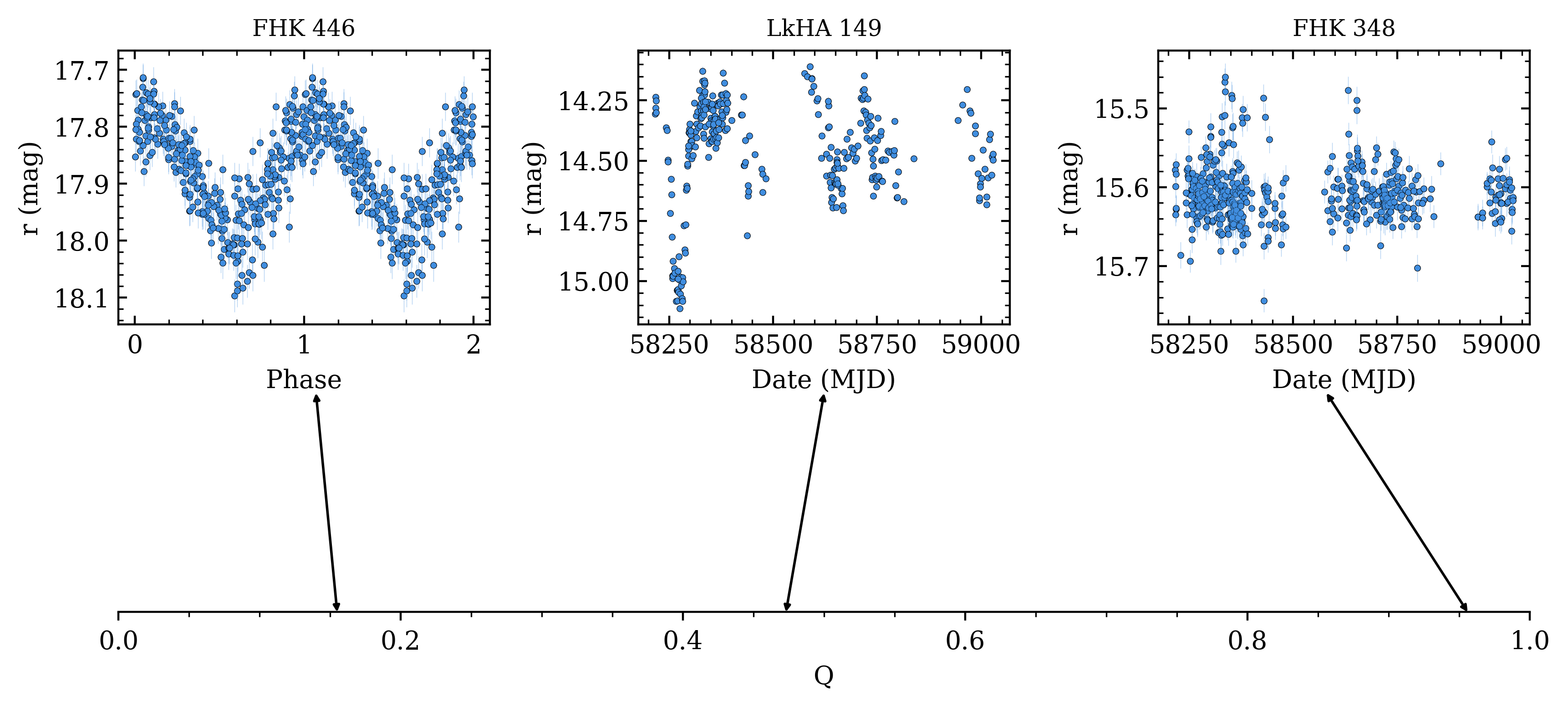}
\centering
\caption{Example ZTF data at low, intermediate, and high values of the asymmetry parameter $M$ (upper panels) and  the quasi-periodicity parameter $Q$ (lower panels).
Negative values of $M$ indicate flaring or bursting type behavior, while positive values of $M$ signify dipping or fading behavior.
Low values of $Q$ indicate more strictly periodic light curves, while high values of $Q$ are characteristic of stochastic light curves.}
\label{fig:qm_tutorial}
\end{figure*}

Our light curve classification scheme follows the methodology of using statistics that quantify 
quasi-periodicity and flux asymmetry, first developed by \cite{Cody2014} and further refined by \cite{Cody2018}. 
Quasi-periodicity, $Q$, is defined as  

\begin{equation}\label{eq:defq}
    Q = \frac{\sigma^{2}_{\mathrm{resid}}-\sigma^{2}_{\mathrm{phot}}}{\sigma^{2}_{m}-\sigma^{2}_\mathrm{phot}},
\end{equation}

\noindent where $\sigma_{\mathrm{phot}}$ is the measurement uncertainty, $\sigma^{2}_{m}$ is the variance of the original light curve, and $\sigma^{2}_{\mathrm{resid}}$ is the variance of the residual light curve after the smoothed dominant periodic signal has been subtracted.
We take $\sigma_{phot}$ as the mean photometric error for all observations in an object's light curve, multiplied by a factor of 1.25 to account for an initial compression of Q values between 0 and $\sim0.6$ that we noticed early in our investigation. We attribute this compression to two effects: 1) the likely underestimate of reported photometric errors, and 2) the cadence dependence of the Q metric, which we discuss in Appendix C. For $\sigma^{2}_{\mathrm{resid}}$, we calculate the residual light curve by adopting a modified version of the periodicity routine described in Section \ref{sec:periodic}. The light curve is first phased to the best period. 
%with the highest power, after potentially aliased periods have been rejected.  
Then, following \cite{Bredall2020}, we mitigate for edge effects by concatenating three cycles of the period and convolving with a boxcar window of width $25\%$ of the period, effectively smoothing the light curve by a factor of four. Finally, we subtract the middle cycle of this model light curve from the folded light curve and compute $\sigma_{\mathrm{resid}}$ as the standard deviation of this residual curve. 
The resulting $Q$ values are expected to fall between $0$ (strong periodicity with low residual noise) and $1$ (completely stochastic behavior). 

Flux asymmetry, $M$, is defined as 

\begin{equation}\label{eq:defm}
    M = \frac{\langle m_{10\%}
    \rangle-m_{\mathrm{med}}}{\sigma_{m}}
\end{equation}

\noindent  where $\langle m_{10\%}\rangle$ is the mean of all magnitude measurements in the top and bottom deciles of the light curve, $m_{\mathrm{med}}$ is the median magnitude measurement, and $\sigma_{m}$ is the standard deviation of the light curve \citep{Cody2018}. Objects that have $M$ values $<0$ are predominately brightening, objects with $M$ values $>0$ are predominantly dimming, and objects with M values near $0$ have symmetric light curves.  In practice, we follow previous literature and assign $M<-0.25$ as bursters, $M>0.25$ as dippers, and the intermediate $M$ values as relatively symmetric lightcurves.

We calculate $Q$ and $M$ based on the $r$-band light curve data.  Figure~\ref{fig:qm_tutorial} provides illustrative light curves along the $Q$ and $M$ sequences.

We then distinguish YSO light curves as populating nine categories, following \cite{Cody2014}. We present these classification results in Table \ref{tab:properties}. 
%These variability classes are defined in depth by \cite{Cody2014}, but we provide brief summaries (with definitions for their notation used elsewhere in this paper given in parenthesis) here for reference (as well as for clarification when different boundaries and/or definitions are adopted). 
In general, the objects are classified based on their $Q$ and $M$ values and then validated by visual examination, but there are cases where we adjust the classification to defer to the human eye for definitive categorization. This is especially prevalent in cases near the morphological boundaries in the $Q-M$ plane. 

Periodic variables (P) are defined as objects with significant periods from Section \ref{sec:periodic} that have $-0.25<M<0.25$ and $Q<0.45$. We discuss our rationale for using a higher $Q$ boundary for periodic objects than adopted in previous publications \citep[e.g.][]{Cody2018, Bredall2020}, in Section \ref{sec:q}.  The variability of the strictly periodic objects is most commonly interpreted as driven by rotational modulation due to the presence of star spots, and the periods derived are thus considered measures of stellar rotation \citep[e.g.][]{Rebull2008}.  
Notable in this regard is that the sources with $Q<0.45$ populate only the lowest range of the normalized variability amplitude,
with $\nu < 0.02$, while sources with higher $Q$ span the full range of $\nu$.
Multi-periodic (MP) objects exhibit more than one distinct period, that is not a beat or alias.  One common explanation for these is star spot modulation in binary pairs \citep[e.g.][]{Stauffer2018,Maxted2018}. 

Objects with $Q> 0.45$ are designated as quasi-periodic or stochastic.  For the quasi-periodics, the variability origin is perhaps related to stellar rotation or the star-disk interaction region, but not dominated by the starspot modulation signal, which can become partially obfuscated in cases of a strong inner disk. There are two flavors of quasi-periodic variables, depending on $M$.

Quasi-periodic symmetric (QPS) variables are defined as having $-0.25<M<0.25$ and  $0.45<Q<0.87$. The variability of these objects is believed to have two possible sources. The first is a combination of purely periodic spot behavior with longer timescale aperiodic changes (e.g. variable accretion), and the second is a single variability process that is unstable from cycle to cycle \citep{Cody2014}.

Quasi-periodic dipper (QPD) variables are categorized based on $M>0.25$ and  $0.45<Q<0.87$. The variability of these objects is believed to stem from variable extinction caused by time-variable occultation by circumstellar material \citep[e.g.][]{Alencar2010, MoralesCalderon2011, Ansdell2016}.
We do not define a quasi-periodic burster category, as distinct from the stochastic bursters, following previous literature in the field. 

Burster (B) variables are defined as having $M<-0.25$, and correspond to objects  with erratic but discrete accretion bursts that are generally short-lived \citep{Cody2017}. These bursts are distinct from larger amplitude and longer timescale outbursts from EX Lup and FU Ori type sources, which are more eruptive \citep{Hartmann1996,Herbig2008}.

Aperiodic dipper variables (APD) are defined as having $M>0.25$ and $Q>0.87$. These objects experience variable extinction, and it has been proposed by \cite{Turner2014} that their variability stems from inner disk scale-height changes induced by magnetic turbulence. 

Stochastic (S) variables are defined as having $-0.25<M<0.25$ and $Q>0.87$.  They feature non-repeating variability patterns with relatively symmetric excursions around a median brightness. 

Long timescale (L) objects exhibit variability on timescales $\gtrsim100$ day timescales.  They generally have $Q>0.87$. 
Finally, we assign $2$ objects to the unclassifiable (U) category because we were unable to calculate their $Q$ due to $\sigma_{\mathrm{phot}}^2$ being larger than $\sigma_{\mathrm{resid}}^2$ and/or $\sigma_{\mathrm{m}}^2$. 

\begin{figure*}[tp]
\includegraphics[]{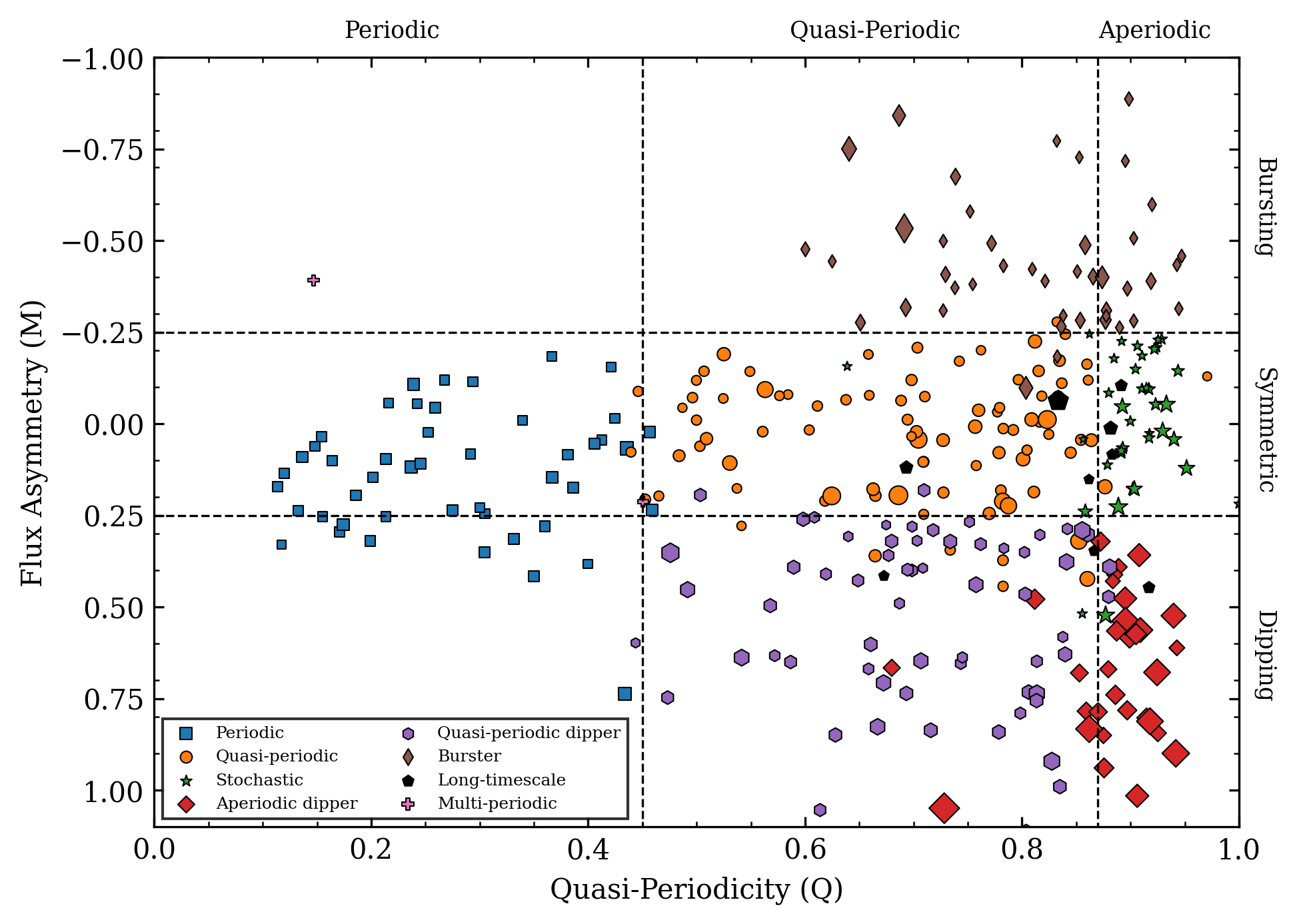}
\centering
\caption{Flux asymmetry versus quasi-periodicity for our sample of variables, color coded by light curve morphological type as reported in Table~\ref{tab:properties}. The size of each point is proportional to its linearly scaled peak-to-peak variability metric $\nu$. Note that some objects, especially periodic sources with high values of $Q$, can fall outside of the defined boundaries between the morphological classes in the $Q-M$ plane.  However, the $Q-M$ classification is fairly robust with only a minor number of sources deserving adjustment to their numerically assigned classes. 
}
\label{qvsm}
\end{figure*}

The results of this classification scheme are reported in Table \ref{tab:properties} for individual sources. Additionally, all stars in the variable sample are plotted in the $Q-M$ plane in Figure \ref{qvsm}, where they are differentiated in color by the assigned (primary) morphological class.    
The distribution of primary and any secondary variability classes is summarized in Table \ref{tab:classesSummary}. 

Objects have secondary classifications in Table \ref{tab:properties} when more than one light curve type is evident in the data. 
Secondary classifications are typically long-timescale (L), representing objects that exhibit a strong primary behavior such as short timescale periodicity, quasi-periodic behavior, or bursting,  along with a pronounced long-term trend such as brightening or dimming, or lasting changes in variability amplitude, etc.   
We call attention to two objects, V1716 Cyg (FHK 437) and FHK 496 
that are designated as periodic in their secondary classification. 
Both have robust periods that phase very well, but their $Q$ values are high, and the $M$ metric reveals that the full amplitude of the light curve is caused by bursty behavior.  
As these cases are each somewhat interesting, we detail them here.

For V1716 Cyg,  %$\mathrm{GDR1}\mathrm{\;} 2162127974652045056$ 
the behavior over most of the light curve is plainly periodic, but for a little over one year of our time series there is superposed continuous bursting at the 1 mag level. \cite{Findeisen2013} also observed bursts in their earlier lightcurve of this source based on PTF data.  
The source further distinguishes itself in having -- aside from the bursts -- a phased lightcurve morphology that appears
(see Appendix A figure set) similar to a W UMa profile\footnote{V1716 Cyg is a fairly secure kinematic and spectroscopic member of the NAP region, with consistent parallax and proper motion, and it exhibits x-ray activity, lithium absorption, and infrared excess.  \cite{Fang2020} reported a spectral type of M1.6 with negligible veiling. W UMa stars are contact or near-contact binaries with two cool components that are both overflowing their Roche lobes.  They are unknown among pre-main sequence stars, but in principle are not implausible given the large pre-main sequence radii, though formation channels would seem to be a challenge.%  
Using the tool %\nolinkurl
\url{https://ccnmtl.github.io/astro-simulations/eclipsing-binary-simulator/} 
we are unable to construct a plausible binary scenario with the expected primary temperature, mass and radius
that produces the observed period, which is a relatively long 4.14 d.  We are thus left with the conclusion that the
phased lightcurve in the non-bursty state is an unusual spot pattern.
%we are able to construct a plausible scenario having 
%$M_1$, $R_1$, $T_1$ = $0.5 M_\odot$, $2.3 R_\odot$, and 3850 K
%and 
%$M_2$, $R_2$, $T_2$ = $0.1 M_\odot$, $2.1 R_\odot$, and 3600 K.
%However, the predicted period for such a system is more consistent 
%with one of the observed period aliases at 1.31 d instead of the 4.14 d we believe to be the true period,
%{\it also the temperature of T2 has to be a bit hot for the mass....} 
}with V-like troughs, and broad inverted-U-like peaks.

In FHK 496, bursting behavior is ubiquitous throughout the light curve, and causes scatter at the 0.4 mag level above the clear periodic signal, which has lower amplitude.

\begin{table}
    \centering
    \caption{Distribution of primary and secondary light curve morphology classifications for objects in our variable sample. Most sources have a single dominant variability type, but $\sim$16\% of the variable sample merited secondary classifications. Values in individual rows have been rounded.}
    \begin{tabular}{c c c}
    \tableline
    \tableline
    Variability Class & Primary $\%$ & Secondary $\%$ \\
    \tableline
    Periodic & 14.6 & 0.9 \\
    Multi-periodic & 0.6 & 1.9 \\
    Quasi-periodic symmetric & 27.6 & 0.3 \\
    Quasi-periodic dipper & 19.2 & - \\
    Aperiodic dipper & 9.6 & - \\
    Burster & 13.9 & - \\
    Long timescale & 2.8 & 12.7 \\
    Stochastic & 11.5 & - \\
    Unclassifiable & 0.3 & - \\
    \tableline
    Total & 100.1 & 15.8 \\
    \tableline
    \end{tabular}
    \label{tab:classesSummary}
\end{table}

\subsection{Color-Magnitude Analysis} \label{colormag}

\begin{figure*}
\includegraphics[scale=0.7]{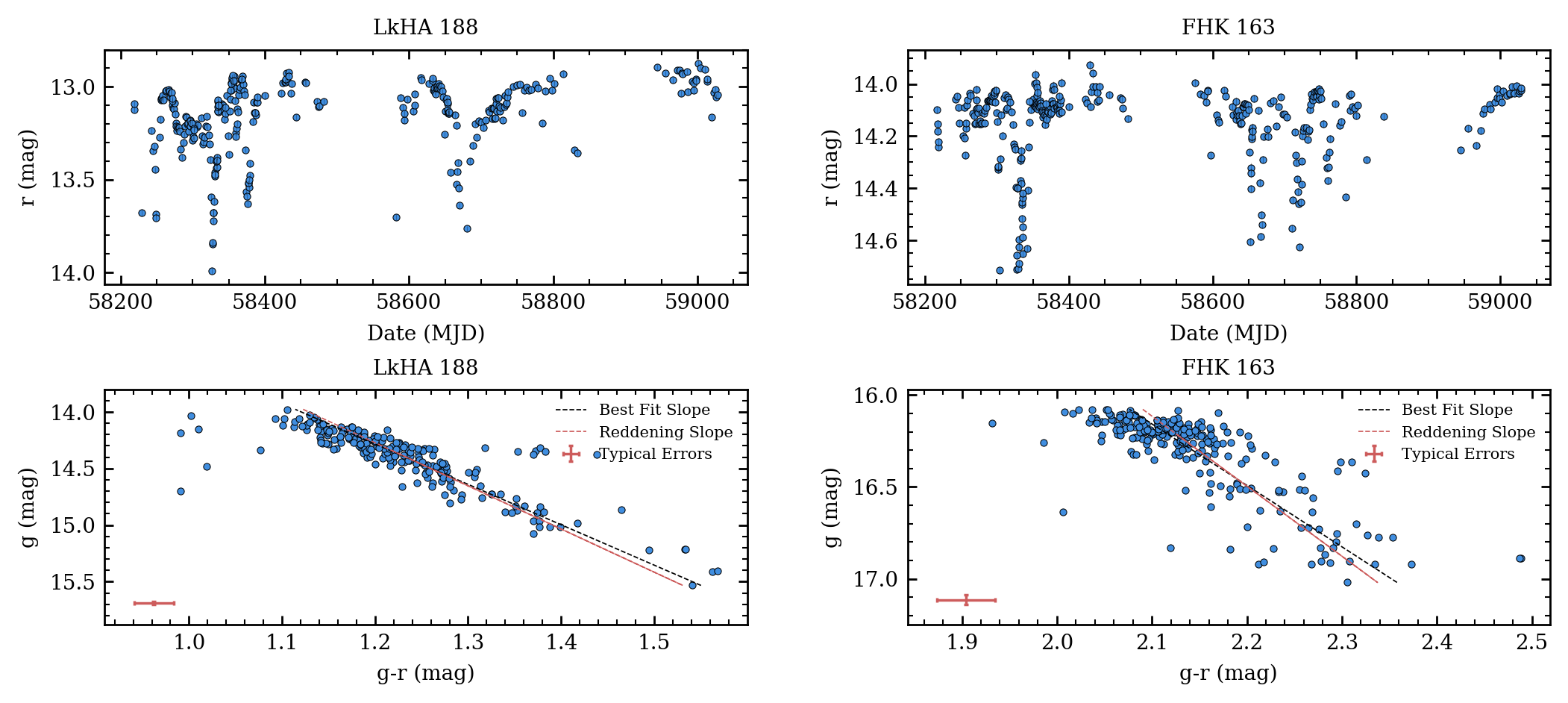}
\centering
\caption{Two exemplary dipper objects with $r$-band light curves shown in the top row, and the corresponding $g$ vs $g-r$ CMDs presented in the bottom row. Both sources have CMD slope angles consistent with their photometric variability being caused by extinction changes, likely along a line of sight that passes through circumstellar material. The expected extinction vector is shown, for reference.}  
\label{slopesGallery}
\end{figure*}

\begin{figure*}
\includegraphics[scale=0.7]{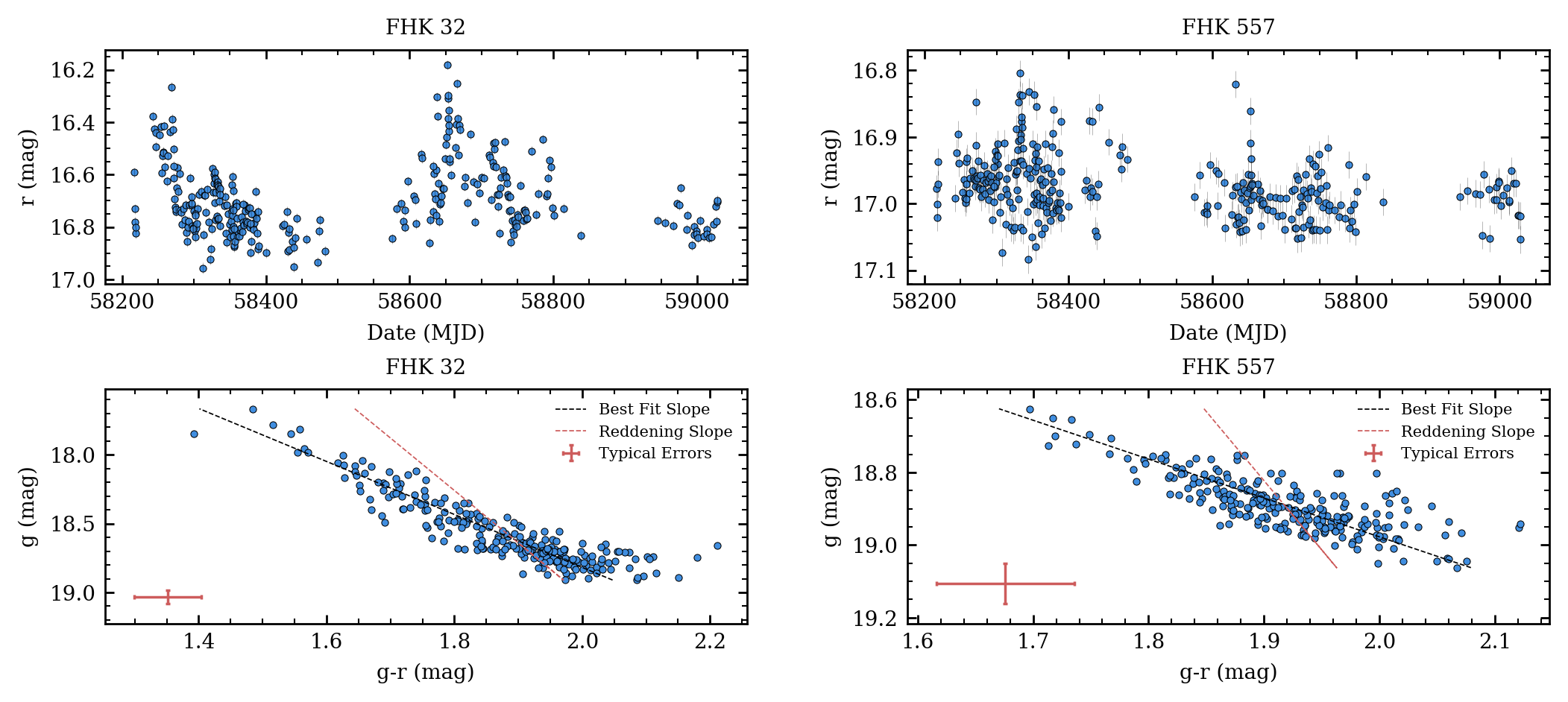}
\centering
\caption{Same as Figure \ref{slopesGallery}, but for two burster objects. Both sources have CMD slope angles much shallower than the expectations from extinction changes. The extinction vector is shown, for reference.}
\label{slopesGallery2}
\end{figure*}

As mentioned above in Section \ref{5.1}, there are many different physical mechanisms that can cause photometric variability in young stars. Color time series data can be a helpful tool in distinguishing them. Among others, \cite{Carpenter2001}, \cite{Guenther2014} and \cite{Wolk2015} showed how near-infrared and mid-infrared multi-color photometry can distinguish extinction-related and accretion-related variability mechanisms.  Optical photometry is even more sensitive to these behaviors, due to the steep rise in extinction from the infrared to the optical, and because of the hot nature of the accretion process. 

The ZTF $g$-band and $r$-band observations are often taken close to one other, within a few hours up to a few days, but they are not simultaneous. Color-magnitude diagrams were thus constructed as follows. First, we partitioned the $g$-band light curves into segments in which the maximum separation between any two subsequent $g$ observations is less than three days.  We adopted this maximum observation separation constraint because linearly interpolating between $g$ observations taken further apart created spurious values in the color-magnitude plane. In practice, within the defined $<$3.0 day intervals, approximately  20.5\% of the $g$-band observations occur within 0.25 days of the prior $g$-band observation, 87.0\% within 1.25 days, 93.4\% within 2.0 days, and 97.5\% occur within 2.25 days.
For each $g$-band value that was paired with an $r$ band observation, we then linearly interpolated between the date bounds of the defined $g$-band intervals, in order to estimate $g-r$ colors. To account for the propagation of errors throughout this process, we employed the Python package \texttt{uncertainties}, which provided the properly calculated errors on the interpolated $g$-band magnitude as well as $g-r$ color values \citep{uncertainties}. 

The time series color-magnitude diagrams exhibit a variety of morphologies, with some sources showing large color and magnitude excursions while others have little color variation, and are essentially constant within the errors.
Figures~\ref{slopesGallery} and ~\ref{slopesGallery2} show examples from among the dipper ($M>0.25$) and burster ($M<-0.25$) categories, which have distinct slope angles.
A selection of sources representing different $Q$ values 
is presented in Appendix A, where a range of CMD slope angles 
can also be seen. Lightcurves and CMDs for the full sample are available in an associated online figure set.

For each member of the variable star set, we performed a linear fit to its color-magnitude diagram (CMD). 
To do so, we applied least-squares linear orthogonal fits using the python package \texttt{scipy.odr} 
\citep{boggs1990orthogonal,Virtanen2020}. This method was chosen following \cite{Poppenhaeger2015},
because it can account for the significant and partially correlated errors in both axes. 
The regression assumes that points lie along a line in the $g$ vs. $g-r$ plane, with Gaussian errors perpendicular to the line. While the condition is not strictly true, the assumption is reasonable given that there is no clear independent and dependent variable in this situation. % and this model describes the expected behavior of extinction. 
To alleviate the effect of outliers, we performed the fits on the middle 95\% of the CMD spans in $g$ and $g-r$. 

After computing the best fit slopes in $g$ vs $g-r$, we define slope angles as the inverse tangent, with the angles in degrees increasing clockwise from $0\degree$ (corresponding to color variability with no associated $g$-band variability) to $90\degree$ (colorless $g$-band variability). To compute errors on the best fit angles, we adopt the \texttt{pYSOVAR} \citep{Guenther2014,Poppenhaeger2015} function \texttt{fit\_twocolor\_odr} and consider slope angles with errors less than $10\degree$ to be significant.  These values are reported in Table~\ref{tab:properties}.
We note that a few sources even exceed $90\degree$ slopes: FHK 101 at $93\degree$,  2MASS J21004676+4255265. %GDR1 2161843098061857280 
at $95\degree$,  and FHK 473 at $97.3\degree$.  
Each of these sources is a large-amplitude variable, with a Pythagorean vector length in the range of 2-3,
and high $Q$ as well as $M<0$.

\begin{figure}[ht]
\includegraphics[width=0.49\textwidth]{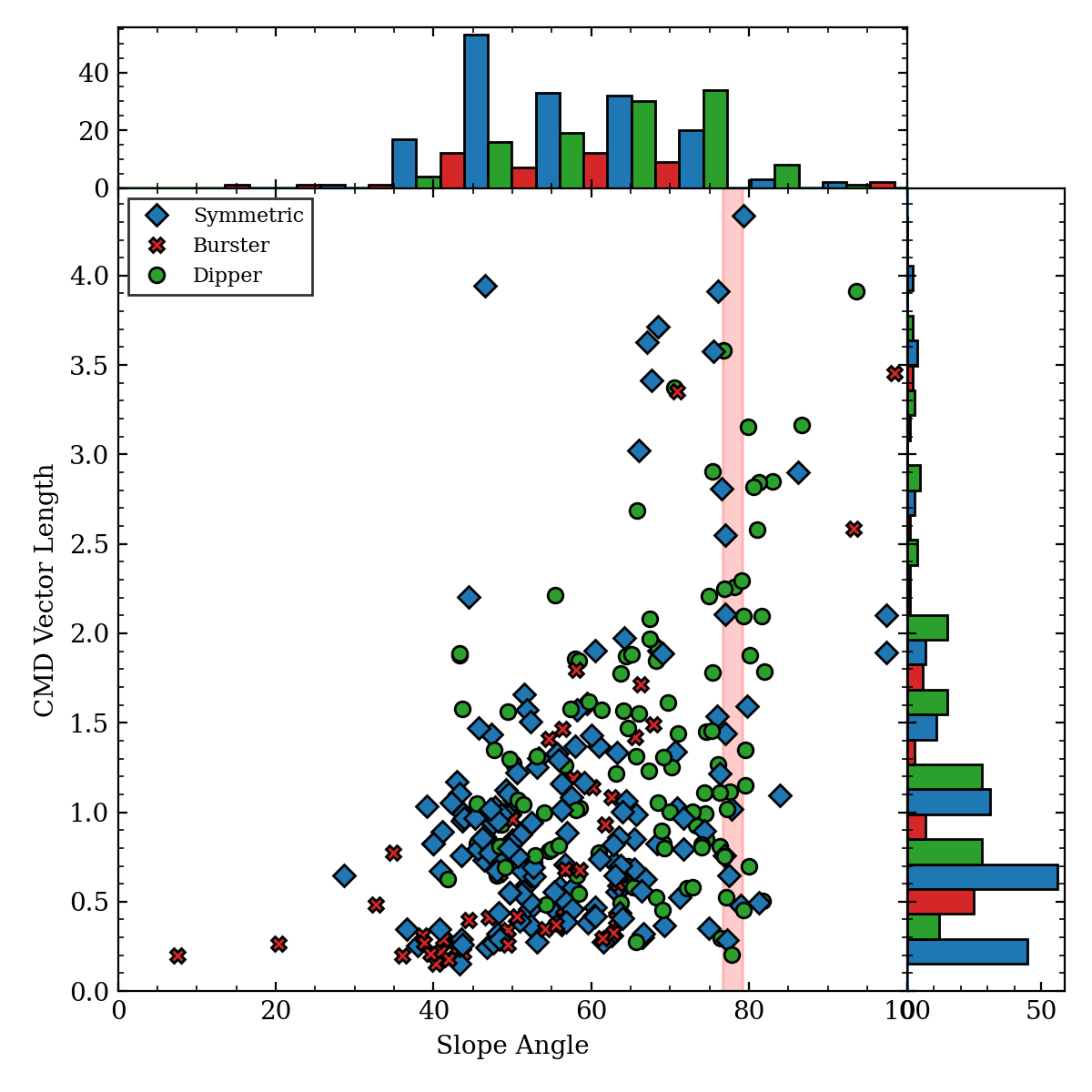}
\centering
\caption{Fitted CMD slope angle in degrees versus Pythagorean vector length in magnitudes, differentiated by flux asymmetry ($M$) class.  The vast majority of sources have slope angles much shallower than the expectations from reddening (shaded band), indicating that the observed color variability is dominated by accretion effects rather than by (or in addition to) extinction effects. 
}
\label{vectorlengths}
\end{figure}

In Figure \ref{vectorlengths}, all significant CMD slope angles are plotted against their CMD vector lengths, following \cite{Poppenhaeger2015}, and differentiated by their flux asymmetry ($M$) classes. There is a sizable span in the CMD vector lengths, with the longest spans limited to the largest slope angles (indicating the largest amplitude and color changes). A wide range of CMD slope angles is also represented, spanning over $45\degree$.   
The outlier at small slope angle is FHK 280,   %GDR1 2162932993257907584, 
which varies over $\sim0.2$ mag in $g-r$ color with very little variation in $g$ brightness; 
in $r-$band it is a low-level burster but we caution that the results could be spurious as the source is near the ZTF bright limit.
The outlier with vector length $\sim 3$ is FHK 332, 
which varies over 2 mag in $g-r$ color and 3 mag in $g$ brightness; it is classified as quasi-periodic symmetric.  
Finally, the extreme vector length $\sim$4 source is FHK 238, an aperiodic dipper.
We also note FHK 267, %GDR1 2162872928138757248, 
the out-of-family burster with vector length $\sim$3.5; this source was also featured in Figure~\ref{fig:qm_tutorial}.

Small slopes, $<35\degree$, are essentially unpopulated, implying a lack of colorless variability.  While likely an astrophysical reality, this could also indicate a bias in our slope determination methods. For example, stars with little to no statistical correlation between variability in magnitude, and variability in color, generally have uncertain slope angles. Such sources are not included in the plot.  However, the full set of slope and vector length determinations includes only 4 (1.2\%)  more sources than the set with $<10\degree$ error that we are plotting. 

The entirety of the populated parameter space in  Figure \ref{vectorlengths} is represented by photometric variables that are 
relatively symmetric in flux (blue). 
The dipper-type variables (green) tend to dominate the population at the longest vector lengths and highest slope angles; their 
median slope angle is $73.8\degree$. 
The burster-type variables (red), conversely, are concentrated towards the shorter vector lengths and populate mainly the lower slope angles, almost exclusively $<70\degree$, with median slope angle $47.7\degree$. 
Beyond just median values, the differences between the morphological dipper and burster classes, and dipper and symmetric classes are very statistically significant based on pairwise K-S tests between their slope angle distributions, with $p<<10^{-11}$. The burster and symmetric difference is also statistically significant, though only with $p=0.013$.

We can assess the CMD slope angles in the context of standard reddening due to dust.
We adopt the extinction curve of \cite{Fitzpatrick1999}, as translated by \cite{Schlafly2011} into the PanSTARRS photometric system (to which ZTF data are calibrated). The expected slopes in $g$ vs. $g-r$ CMDs for different reddening laws are: 3.52 for $R_V=3.1$, 4.39 for $R_V=4.1$, and 5.24 for $R_V=5.1$, with steeper slopes corresponding to larger dust grain sizes, as might be expected in molecular clouds and circumstellar disks.  Considering a range of possibilities for the extinction law, we expect objects with time-variable CMD behavior dominated by extinction effects to have slope angles between $74.1\degree$ and $79.2 \degree$ in the $g$ vs $g-r$ diagram.  

Examination of Figure \ref{vectorlengths} shows, however, that the majority of slopes are much shallower than the expectations from purely reddening variations. 
We interpret the distribution of slope angles as indicating that much of the observed variability is at least partially related to accretion effects, with extinction effects (likely occurring outside of the accretion zone) perhaps playing some role.  
We acknowledge that other morphological light curve types can have slope angles in the CMD consistent with reddening, that is, not all sources that look like they are experiencing variable extinction also exhibit identifiable dipping behavior.

\section{Correlation of Variability Properties with Infrared Excess} \label{sec:disks}

\begin{figure}[htp]
\includegraphics[width=0.49\textwidth]{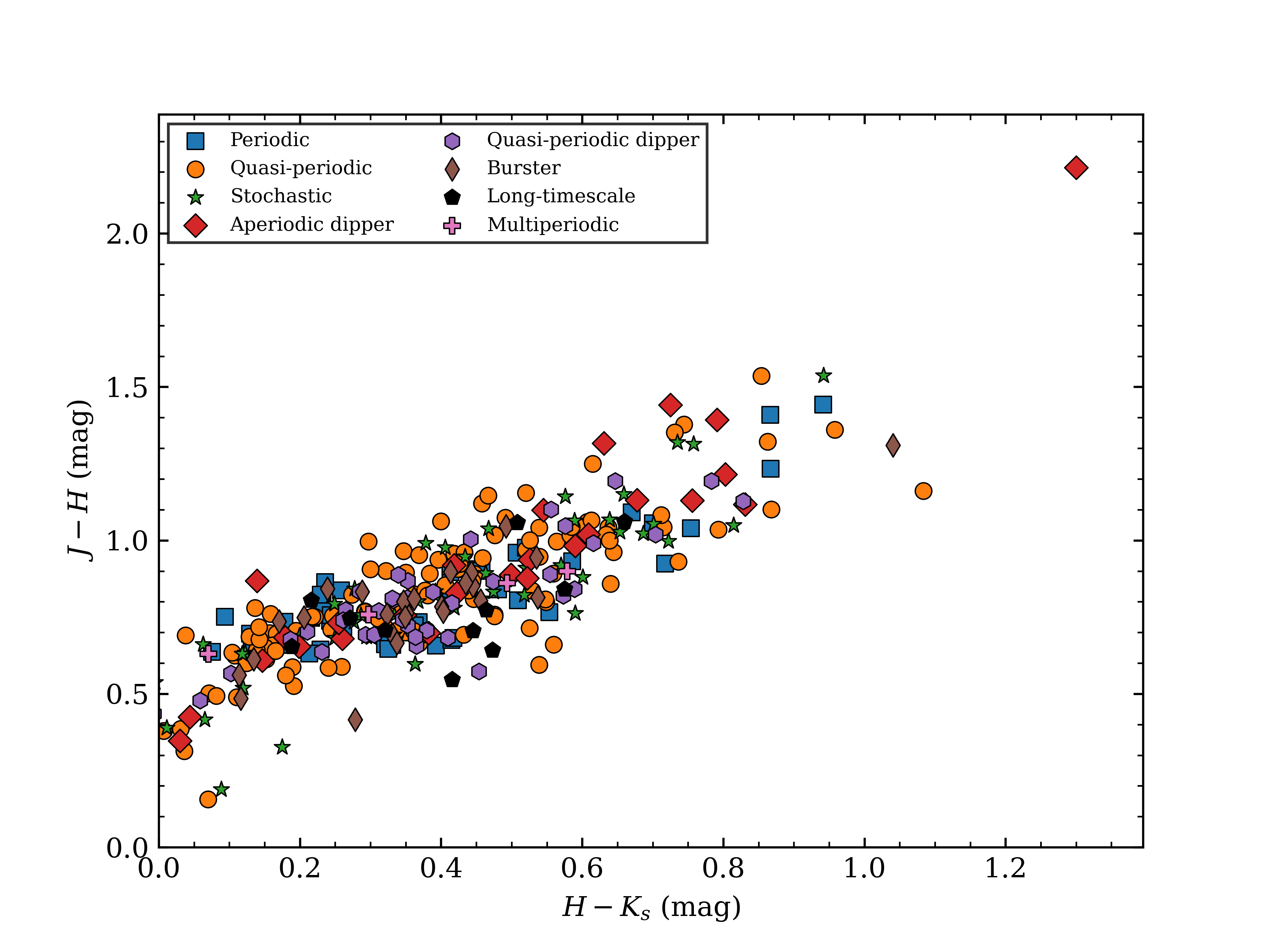}
\includegraphics[width=0.49\textwidth]{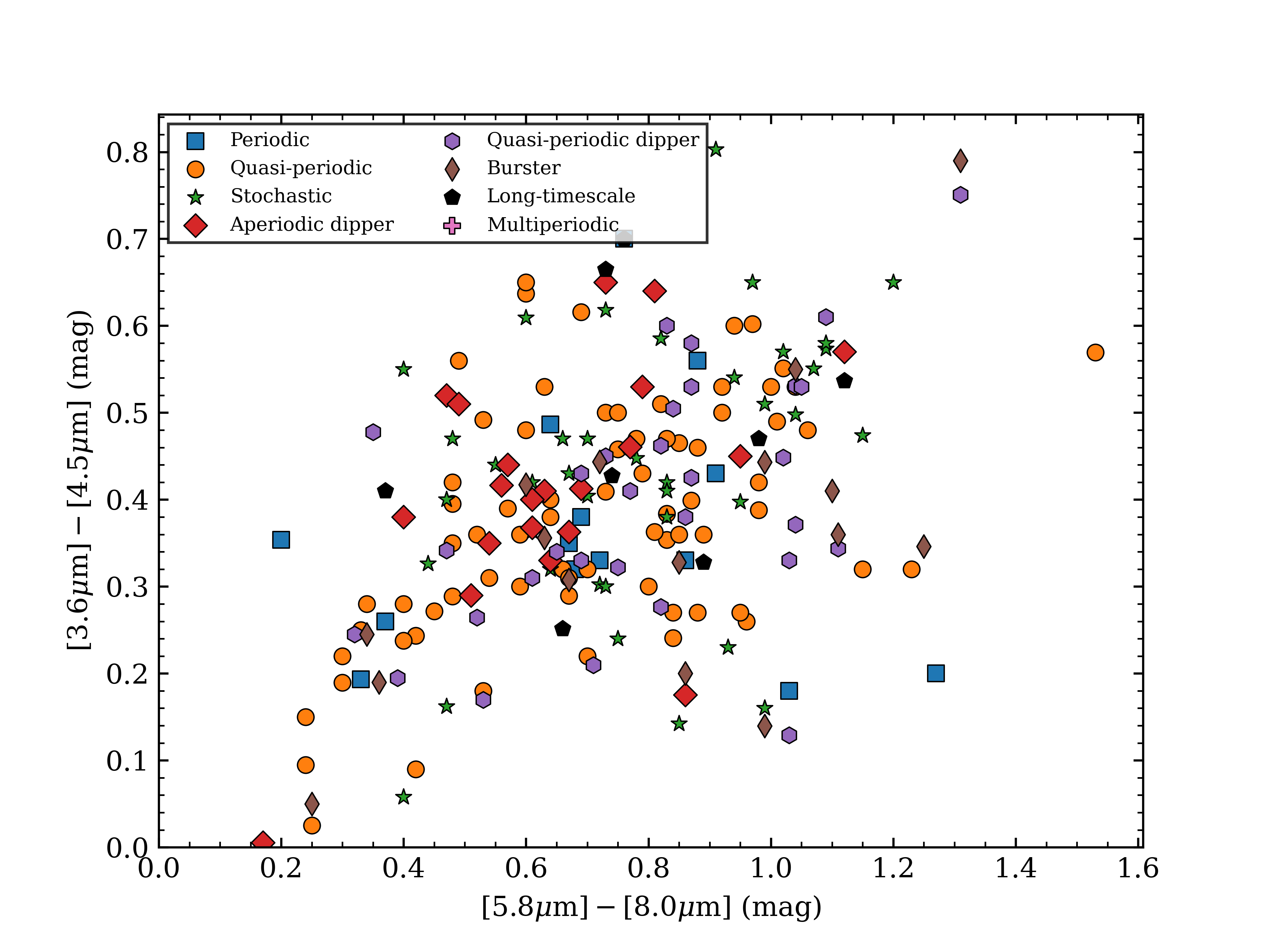}
\includegraphics[width=0.49\textwidth]{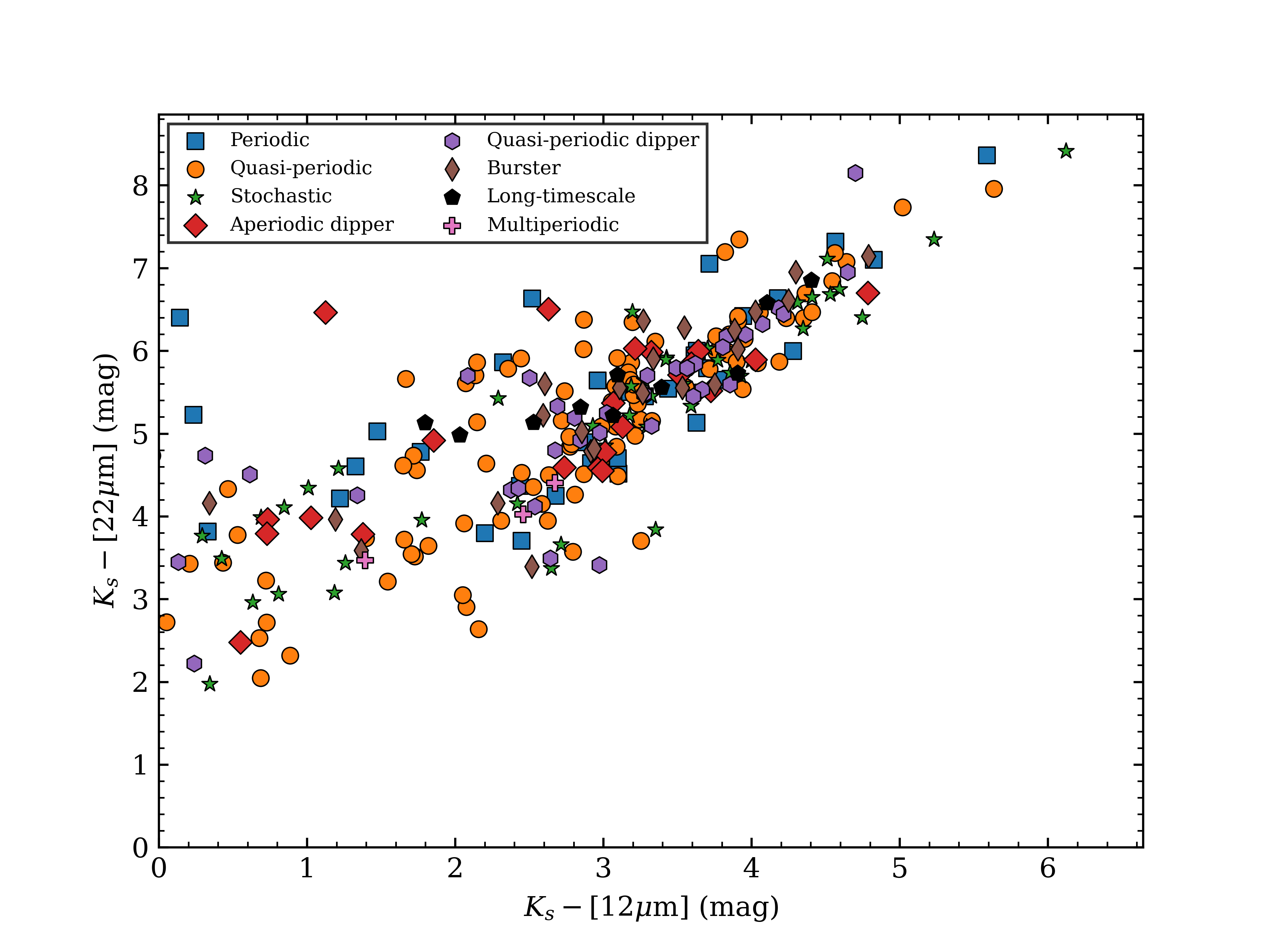}
\caption{Dereddened infrared color-color diagrams with objects having different $Q-M$ classifications distinguished. Except for the bluest 5-10\% of the sample, the sources have colors consistent with those of Class II YSOs.  The objects classified with periodic light curves tend to have small infrared colors, while those with other behavior tend to have larger infrared colors, indicative of excesses consistent with the presence of circumstellar dust.}
\label{fig:infraredcolor}
\end{figure}

\begin{figure}[htp]
\includegraphics[width=0.49\textwidth]{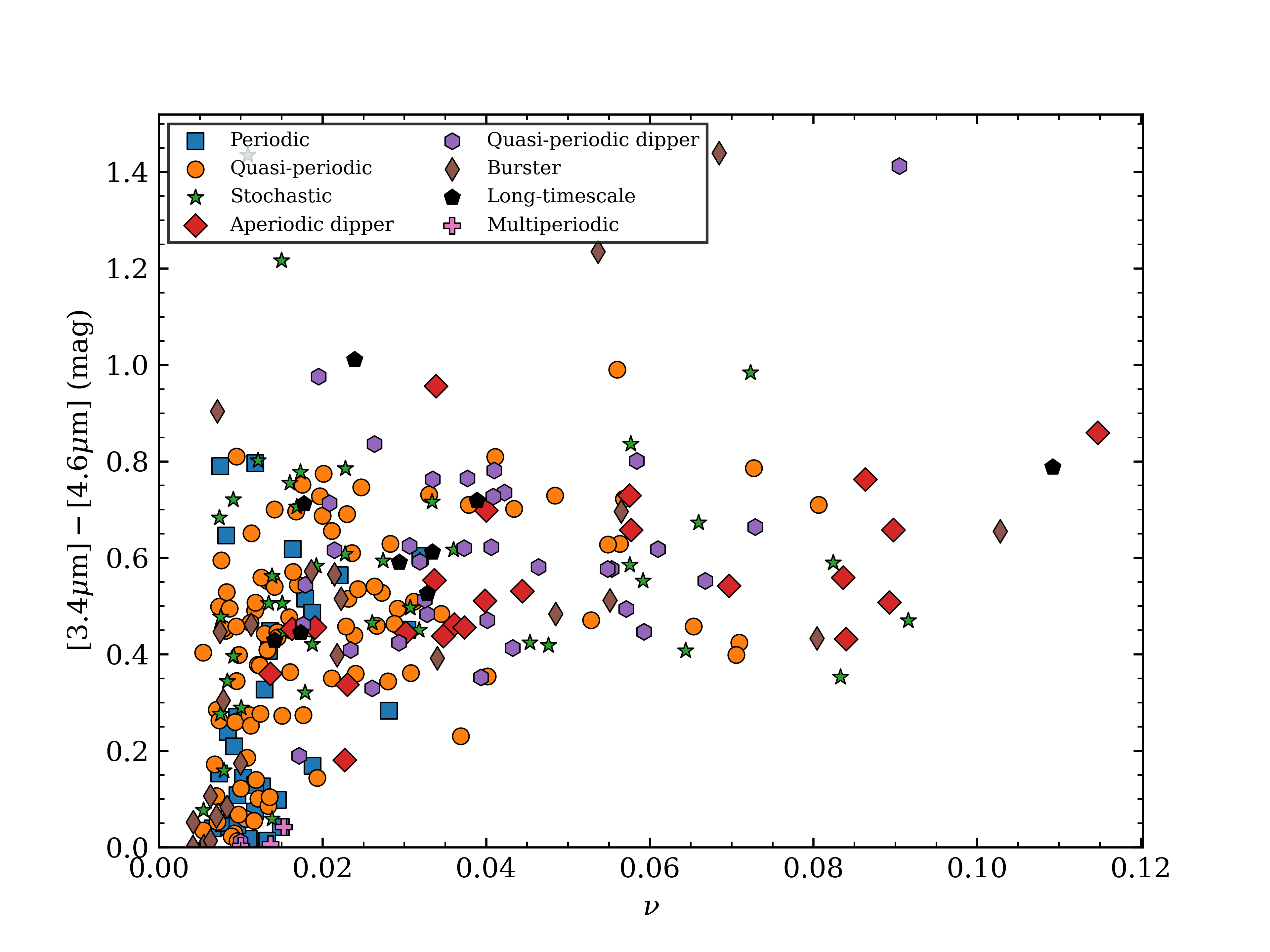}
\centering
\caption{Dereddened infrared color vs $\nu$ metric, measuring normalized amplitude of the optical variability. 
Colors redder than 0.2 mag in these bandpasses indicate young stars with infrared excess.  
Essentially all sources with $\nu > 0.02$ appear to have excess infrared color, indicating young stars surrounded by circumstellar dust.}
\label{fig:infrarednu}
\end{figure}

There is some expectation that the light curve morphological classes could correlate with infrared excess.  Using $K2$ data sets,
several previous studies, namely \cite{Cody2018} in Oph/Sco, \cite{Cody2022} in Taurus, 
and \cite{Venuti2021} in the Lagoon Nebula all found that both the periodic (P) 
and the quasi-periodic symmetric (QPS) sources tend to have lower values of infrared excess on average,
and in fact dominate the source population at small infrared color values.  
This is consistent with a relatively clean line of sight to the spotted stellar photosphere, 
and a lack of accretion effects in the light curve.  
Sources with higher $Q$, indicating quasi-periodic and stochastic light curves, 
as well as the dippers and bursters, on the other hand,  had larger values of infrared excess in these previous studies.

To explore whether these correlations can be seen using light curve morphologies assigned 
based on ZTF data, we cross-matched our variable star sample with photometry from
2MASS \citep{Cutri2003}, WISE \citep{Wright2010,Cutri2012}, and Spitzer \citep{Rebull2011}.
Figure~\ref{fig:infraredcolor} shows several color-color diagrams with the light curve morphologies 
from Table~\ref{tab:properties} distinguished. 
Figure~\ref{fig:infrarednu} shows one infrared color as a function of our normalized variability amplitude metric.
Relative to the variability morphology vs infrared color analysis in \cite{Cody2018} studying Upper Sco and \cite{Cody2022} studying Taurus, we may be more subject here in the NAP region to the effects of reddening.
Thus, to de-redden the color-color diagrams, we adopted the $A_V$ values of in \cite{Fang2020} 
and the near- and mid-infrared extinction results  of \cite{xue2016}. 
Among our optical variables, about 78\% have extinction values reported in \cite{Fang2020}.
For the remainder, we adopted the median value of $A_V$.

Previously identified (weak) trends 
between light curve morphology and infrared excess are not clearly present in our analysis.
We note that our ground-based data has lower cadence and precision than $K2$ data, 
and that we are thus less sensitive to the smaller-scale variability patterns 
that can be discerned in the exquisite $K2$ light curves.

%%%%%DISCUSSION%%%%%%%%
\section{Discussion} \label{sec:discussion}

Our discussion focuses on three aspects of our analysis.
In \S\ref{sec:periodrecov} we compare our periods to those found 
in previous literature.  In  \S\ref{sec:q}  
we philosophize about the broad applicability of the quasi-periodicity $Q$ metric, in particular regarding the need to tailor its boundaries for use in ground-based data sets.   
Finally in \S\ref{sec:qmcompare}, we assess how the 
distribution of sources in the $Q-M$ plane for the NAP region compares to what has been found in other star forming regions.

\subsection{Period Recovery for Strictly Periodic Variables}\label{sec:periodrecov}

% from Rebull: it feels like this should go earlier in the paper, next to "we derived periods this way". it make sense to me to follow this by "and we can do so as well as the literature."

\begin{figure}[htp]
    \centering
    \includegraphics[width=0.4\textwidth]{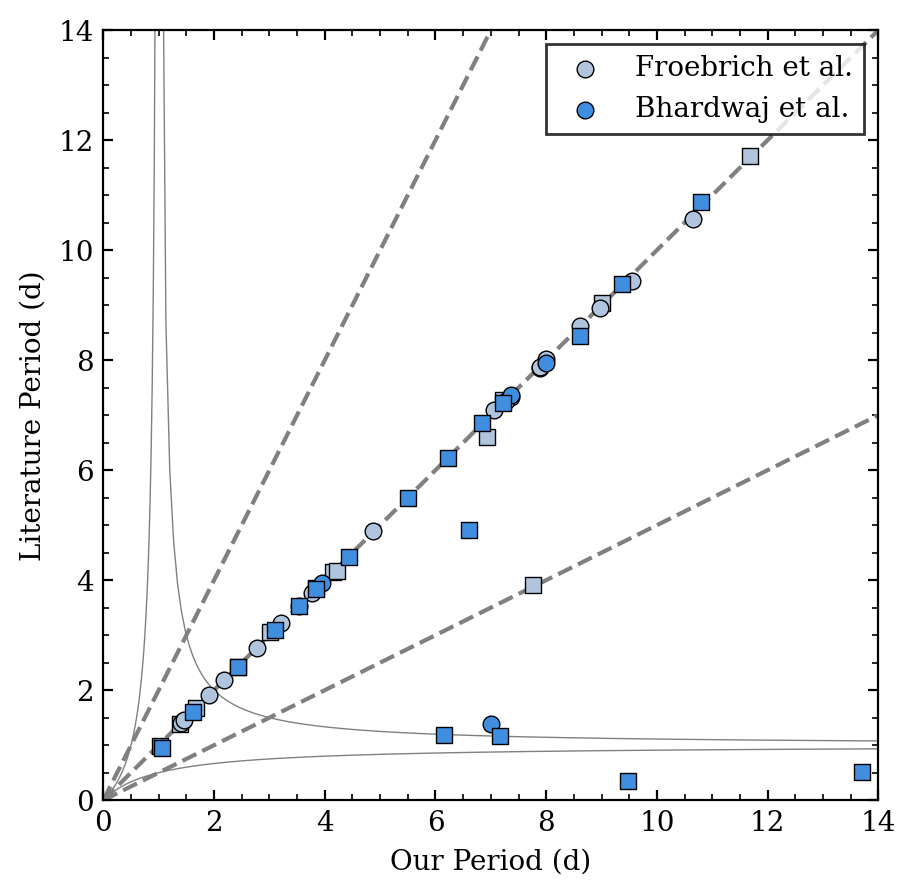}
    \caption{Comparison of periods for objects that overlap between our work and previous literature.  Lines indicate the 1:1 relationship between the periods P, and the P/2 and 2P harmonics (dashed), and the $(|1/P \pm 1)|^{-1}$ beats with the $\sim$1 day sampling (dotted).  Circles are sources we find to exhibit true periodic behavior, whereas squares are sources that we designate as having timescales rather than strict periods, based on their high $Q$ values.} 
    \label{fig:periodRecovery}
\end{figure}

In order to test the quality of our derived periods, we compared them to the periods published in \cite{Bhardwaj2019} and \cite{Froebrich2021}. Both papers considered objects in the Pelican Nebula region (see Figure~\ref{fig:skyview}). 

Of the 40 periodic variables identified in \cite{Froebrich2021}, $32$ are within $2$\arcsec\ of a source in our sample, of which 20 we classify as periodic in $Q-M$ space. All 20 have periods  within $5\%$ of our derived periods. Additionally, among the 56 objects found to have strong periodic variations in \cite{Bhardwaj2019}, 27 variables match within 2\arcsec\ of a source in our sample and we classify $4$ as periodic in $Q-M$ space.
In this sub-sample, we recover $3$ periods within $5\%$ of their \cite{Bhardwaj2019} values. We compared the phase-folded light curve of the remaining object with the discrepant period and found that forcing it to the period in \cite{Bhardwaj2019} introduces pronounced additional scatter.   % or provides no improvement in phase dispersion. 
Among the additional $23$ stars in common, $18$ have periods in agreement within 5\%, though we have labelled these periodogram peaks as timescales in Table~\ref{tab:properties} rather than periods, due to poor phasing and/or high $Q$ values.

Overall, between these two samples our period recovery rate for strictly periodic objects is $95.8\%$. These period comparisons are presented in Figure \ref{fig:periodRecovery}. 

%The two outliers and their contrasting periods are: 
%ID,Our Per,Bhard Per
%2MASS_J20510029+4424364,7.008938672886845,1.39
%FHK_26,13.704636645102322,0.518

\subsection{Comments on the Quasi-Periodicity Metric $Q$} \label{sec:q}
The quasi-periodicity metric Q has proven to be an extremely powerful tool for classifying the variable behaviour of YSOs \citep{Cody2014, Cody2018}. In this work, we have shown that the metric is sufficiently applicable to ZTF data, affirming its utility for ground-based as well as space-based photometric data sets. However, recent studies have revealed issues in translating Q results between various sources of data \citep[][]{Cody2018, Bredall2020}.  

When comparing objects studied with different space-based telescopes, systematic effects present in the precision photometry from different observatories, or those caused by slight wavelength dependencies of the astrophysical phenomena, could produce systematically different Q values for sources with similar intrinsic behavior \citep{Cody2014, Cody2018}.  Notably, the morphological boundaries in $Q$ were adjusted going from analysis of CoRoT data in \cite{Cody2014} to K2 data in \cite{Cody2018}. For the CoRoT data, \cite{Cody2014} set Q boundaries at 0.11 and 0.61, with M boundaries at $\pm$ 0.25.   For K2 data, in \cite{Cody2018} the adopted Q boundaries were 0.15 and 0.85, with M remaining at $\pm$ 0.25.

In contrast to this adjusting of the $Q$ metric, \cite{Bredall2020} did not adjust the morphological boundaries of $Q$ in their ground based study based on ASAS-SN data, instead opting to simply apply the morphological boundaries defined for K2 data in \cite{Cody2018}.   These authors did note the discrepancies that appeared in their classification scheme, such as objects having been classified in previous variability studies as QPDs, assessed in their numerical analysis as strictly periodic objects.

Indeed, prior to tailoring our Q routine to the ZTF data set, we noticed a similar compression and shift of our objects towards the periodic end (low $Q$) of the quasi-periodicity scale, with an absence of objects having $Q > 0.65$.   This would be a very different distribution from the results of \cite{Cody2014} and \cite{Cody2018}, as in both of these studies, the range from Q=0 to Q=1 is densely populated \citep[in fact, exceeding these bounds in][]{Cody2014}.

Given the substantially larger errors and the lower observational cadence associated with ground-based data, significant discrepancies in $Q$ values might be expected when comparing objects studied with space-based versus ground-based telescopes. 
In Appendix C we investigate in detail the effects that cadence and photometric uncertainty have on $Q$.

For our analysis, we opted to follow the methodology of \cite{Cody2018}, tailoring both our morphological boundaries as well as our procedure for calculating Q to the nuances of our particular ZTF data set. As stated in \S \ref{5.1}, our adopted Q boundaries are  0.45 and 0.87. We selected these particular values after visually inspecting all variable object light curves, both unfolded and phase-folded, and comparing them to light curves from both \cite{Cody2014} and \cite{Cody2018}.  We assigned the $Q$ boundaries to occur at the most obvious behavioral transitions we noticed in the light curves when ordered by $Q$, and to minimize the number of edge-case objects that we end up re-classifying by eye.  After visual inspection, approximately $11\%$ of objects were re-classified. 

We believe that such a flexible treatment of the morphological boundaries indicated by $Q$ is the best way to ensure that variables classified in the $Q-M$ parameter space are behaviorally representative of the original classes as defined by \cite{Cody2014}.  In other words, there is an un-quantified error inherent in the calculation of the $Q$ and $M$ metrics, and it is more useful for furthering our understanding of the origins of young star variability, and for comparing sources observed with different instruments, if the quantitative procedures and morphological boundaries are tailored to the instruments used in each analysis. While objects of the same intrinsic morphological class could appear in different Q ranges in different studies, they should be properly grouped with their light curve peers.  

Regarding the full range spanned by our derived $Q$ values (about 0.11-0.96), we believe there are reasonable explanations why the full zero-to-one parameter space is not fully covered.  On the low side,  none of our objects have $Q$ values between 0.0 and 0.1 partially due to our choice to represent the estimated photometric uncertainty with the mean photometric error for each object, and our further inflation of the photometric errors during the $Q$ calculation process.  This imposes an artificial minimum to $Q$. There is also the likely reality that no objects in our sample have negligible phase dispersion (including photometric uncertainty) relative to the amplitude of a hypothetical periodic oscillation.  On the high side of $Q$, the maximum at 0.96 instead of 1.00 may indicate simply that we have not inflated our errors enough to produce a truly stochastic $Q$ metric.  On the other hand, there is no rationale for fine tuning so as to have any particular number of objects with the theoretical maximum value of $Q=1.00$. 

Regarding the range spanned by the $M$ values (about -1.2-1.1) for our aperiodic objects, there is again no expectation that any given variable star sample should span, or exist constrained within, the theoretical range from -1 to +1; although we observe our results seem to appropriately populate the parameter space in almost all cases. We note that, as the procedure for calculating the flux asymmetry $M$ metric is more agnostic to the instrumental effects than the procedure for calculating $Q$,  we were comfortable adopting the same flux asymmetry boundaries used in previous studies to designate dippers, symmetrics, and bursters. We note that the periodic objects at  $Q<0.45$ occupy a more restricted range in $M$, closer to the symmetry line of $M=0$.   

Summarizing, as these metrics are further applied, we recommend the continued utilization of visual inspection. We further suggest that $Q$ values be scaled between approximately $[0-1]$ in future works, in order to standardize the use of this metric across data sets.

\subsection{Comparison to $Q-M$ Distributions in Other Star Forming Regions}
\label{sec:qmcompare}

In Table~\ref{tab:classesSummary}, we summarized the fraction of objects in the member sample 
we have assembled for the North America and Pelican Nebula region, in the different variability classes. 
We can compare our relative distribution to those resulting from previous uses of the $Q-M$ parameter space
to study young stars. 

To do so, we refer to Table 3 in \cite{Cody2022}
which includes the Taurus, Oph, and NGC 2264 regions studied with \textit{K2} and \textit{CoRoT} data taken from space-based platforms.
We focus our comparisons on the young regions with comparable ages to the NAP. 
The only other region with such information, Upper Sco, has variability properties that are distributed 
somewhat differently, perhaps because Upper Sco is somewhat older.
We do not consider the results of \cite{Bredall2020} from ground-based data in Lupus, due to 
the apparent distortion in their $Q$ distribution towards low values; this may be a sign of the issue we faced in this study
(\S 5.1) where inflating our photometric errors by 25\% recovered more of the expected range in $Q$ for a young star variable sample.
There is also the issue of whether ground-based and space-based data sets are comparable.
As mentioned earlier, we assess the effects of data precision and data cadence on $Q$ in Appendix C.

Relative to the previously studied star forming regions, we find a similar fraction of periodic sources, 
including the periodic (P) and multi-periodic (MP) classes, at around 15\% of the sample.
This is interesting given that our sample is designed to include all members and not just disk-bearing sources
that comprised the samples studied previously. The result can be understood if we accept that our sample 
is more biased towards including periodic sources, but at the same time we are less capable of detecting them as such,
given the lower quality of our data set.
Another commonality is that the quasi-periodic symmetric (QPS) category is our largest, 
and similar to the other star-forming regions at about 28\%.
%our fraction is even higher than in the space-based studies likely due to the noisier nature of our ground-based photometry.
Our stochastic fraction at 11\% also matches well to the previous results.
The quasi-periodic and aperiodic dipper (QPD and APD) groups are also similar, cumulatively around 29\% of our sample.
There is a range in these fractions among the other star-forming regions over a factor of two, with our value in the middle.  
Our burster fraction is 14\%, on the low side of the other regions (which range from about 13-21\%).
An explanation here is that much of the burster activity identified from $K2$ is short-lived 
\citep[e.g.][]{Cody2017} and stands out relative to what could be detected from the ground. 

Summarizing, our ground-based data set has matched the variability properties of space-based studies of comparably-aged stars
rather well, with any differences plausibly attributed to
our lower sensitivity from the ground to narrow bursts and accretion flaring activity.

%%%%%CONCLUSION%%%%%%%
\section{Summary}
\label{sec:conclusion}

In this work we have studied the variability of young stellar objects in the North America and Pelican Nebulae region using multi-color observations from $>800$ days of Zwicky Transient Facility measurements.  Our primary results are as follows: 

\begin{itemize}
    \item[-] We have searched for periodicity in the light curves of all variable objects in our sample and compared our derived periods with those previously published in the literature.
    
    \item[-] We have classified 323 stars using the $Q-M$ variability plane, to quantify flux asymmetry and waveform repeatibility in the light curves.  
    Of these, 44 objects are classic periodic variables and 2 additional sources show bursting behavior on top of very strong periodic behavior; several objects are multi-periodic variables. Among the rest of the sample, there are 88 quasi-periodic symmetric variables, 62 objects quasi-periodic dippers, 31 aperiodic dippers, 46 bursters, 36 stochastic variables, and 9 long-timescale variables. We additionally designated 45 objects with secondary classifications where more than one behavior is exhibited.
    
     \item[-]
     The dominant variability, characterizing $\sim$55\% of the sources, is relatively flux-symmetric behavior 
     with low absolute value on the $M$ index.  Included in this 55\% are the $\sim$15\% of the sample 
     in the strictly periodic regime of $Q$, having variability signals similar to starspot patterns, 
     plus another $\sim$39\%  of the sample with flux-symmetric $M$ values, but quasi-periodic or stochastic $Q$ values.   
     The quasi-periodics are interpreted as sources in which an underlying rotationally driven signal is masked by circumstellar activity
     occurring within or near the stellar co-rotation radius in the inner disk.
     Among the flux-asymmetric sources, having higher absolute value $M$ measurements, 
     the $Q-M$ analysis reveals $\sim$14\% bursters  and $\sim$29\% dippers.
     These fractional distributions are roughly consistent with those observed in other clusters of comparable age and disk fraction.
     
     \item[-]
     The fraction of periodic or quasi-periodic sources, having small and moderate values of $Q<0.87$, is 68\% while
     the fraction of more erratic variables, with larger $Q$, 32\%. 
     This ratio is similar to that in Taurus with 70\% and 30\%, respectively \citep{Cody2022}.

    \item[-] We have compared the light curve morphology classes in terms of their variability in CMDs in order to investigate the physical drivers of variability.  We found a clear distinction in the distribution of CMD slope angles for dippers and bursters, with dippers closer to the expectations from extinction-driven variability phenomena, and bursters exhibiting much flatter slopes that are suggestive of accretion-related variability drivers.

    \item[-] We have investigated and discussed various subtleties of the quasi-periodicity metric $Q$, and recommend methods for its use in future studies. Specifically, as acknowledged in previous literature, the metric is sensitive to the accuracy of the photometric errors 
    and further, different data sets necessitate flexible treatment of the light curve morphology boundaries using the $Q$ metric.  We have also shown that, while photometric uncertainty is important, and the lower precision of ground-based data causes us to miss some low-level variability that is detectable in data from K2, the lower cadence of ZTF has a more pronounced effect on $Q$, substantially reducing it for certain categories of variability.

    \item[-] The python code used to calculate the $Q$ and $M$ variability metrics is available at 
    \url{https://github.com/HarritonResearchLab/NAPYSOs/tree/main/results}.
    %\url{https://github.com/HarritonResearchLab/NAPYSOs/results}.
    %https://github.com/HarritonResearchLab/NAPYSOs/blob/main/results/functions.py
\end{itemize}

\section{Acknowledgements}
%\begin{acknowledgements}
We thank Leah Seignourel, Pietro Romussi, and Korgan Atillasoy for their participation in early discussions about this work. We also thank John Bredall, Katja Poppenhaeger, and Hans Moritz Günther for assistance with various Python questions, as well as Travis Austin for assistance with recovering a significant amount of our work from a damaged virtual machine disk. Ann Marie Cody provided the K2 light curves that form the basis of Appendix C, as well as valuable advice concerning $Q$. This work is based on data from the Zwicky Transient Facility, which is supported by the National Science Foundation under Grant No. AST-1440341 and a collaboration including Caltech, IPAC, the Weizmann Institute for Science, the Oskar Klein Center at Stockholm University, the University of Maryland, the University of Washington, Deutsches ElektronenSynchrotron and Humboldt University, Los Alamos National Laboratories, the TANGO Consortium of Taiwan, the University of Wisconsin at Milwaukee, and Lawrence Berkeley National Laboratories.
We thank the referee for a careful look at our methods and results.
%\end{acknowledgements}

\facilities{P48:ZTF, IRSA}

\textit{Software:} NumPy \citep{numpy}, Matplotlib \citep{Hunter2007}, Pandas \citep{pandas}, Astropy (\cite{astropy1}, \cite{astropy2}), SciPy \citep{Virtanen2020}, and uncertainties \citep{uncertainties}. 

%\newpage
%\vfill\eject

%%%%%%REFERENCES%%%%%%%
%\bibliographystyle{apj} 
\bibliography{refs.bib}

\begin{appendix}

\section{Individual Object light curves, Phased light curves, and Color-Magnitude Diagrams}

\begin{figure*}[tp]
\includegraphics[width=0.7\textwidth]{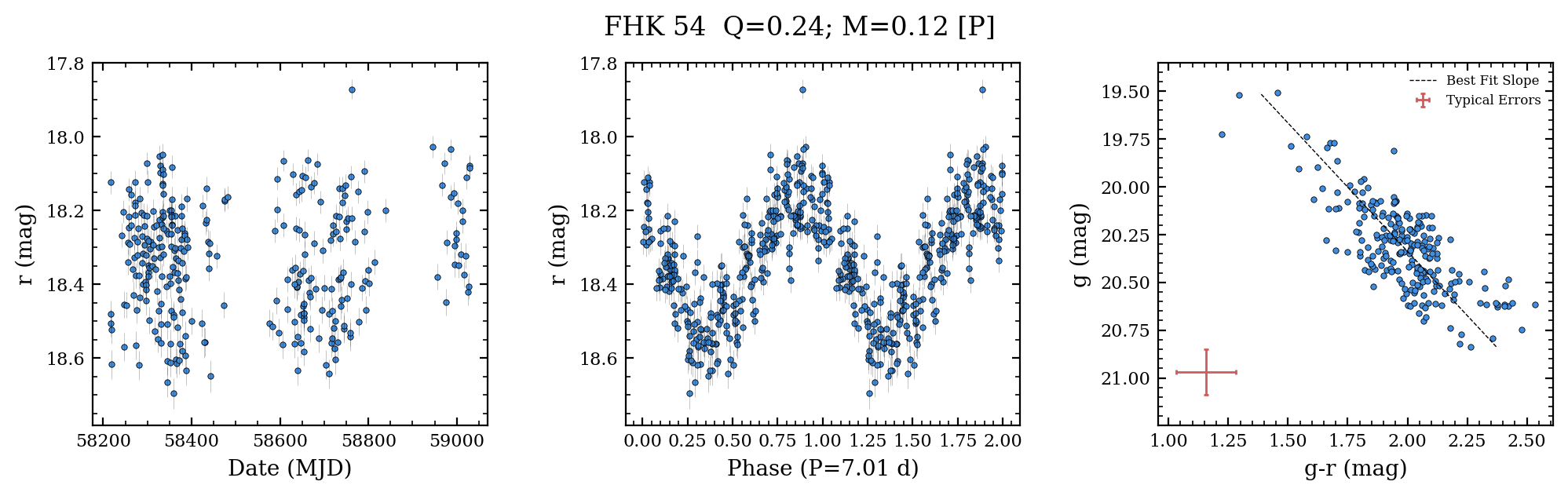}
\includegraphics[width=0.7\textwidth]{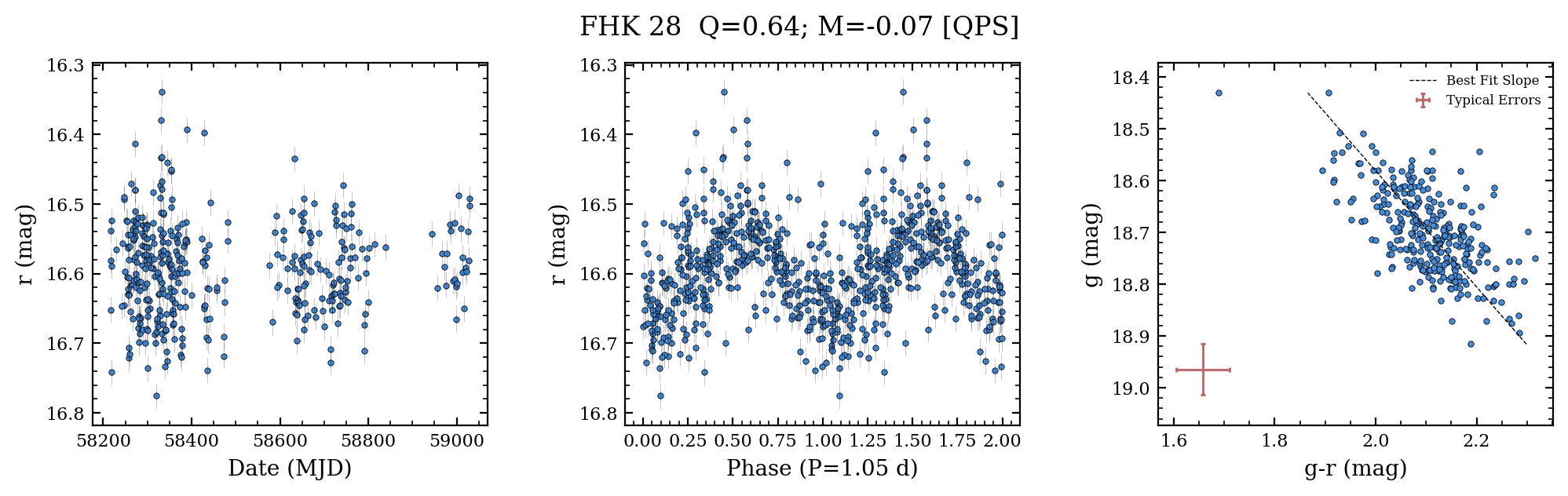}
\includegraphics[width=0.7\textwidth]{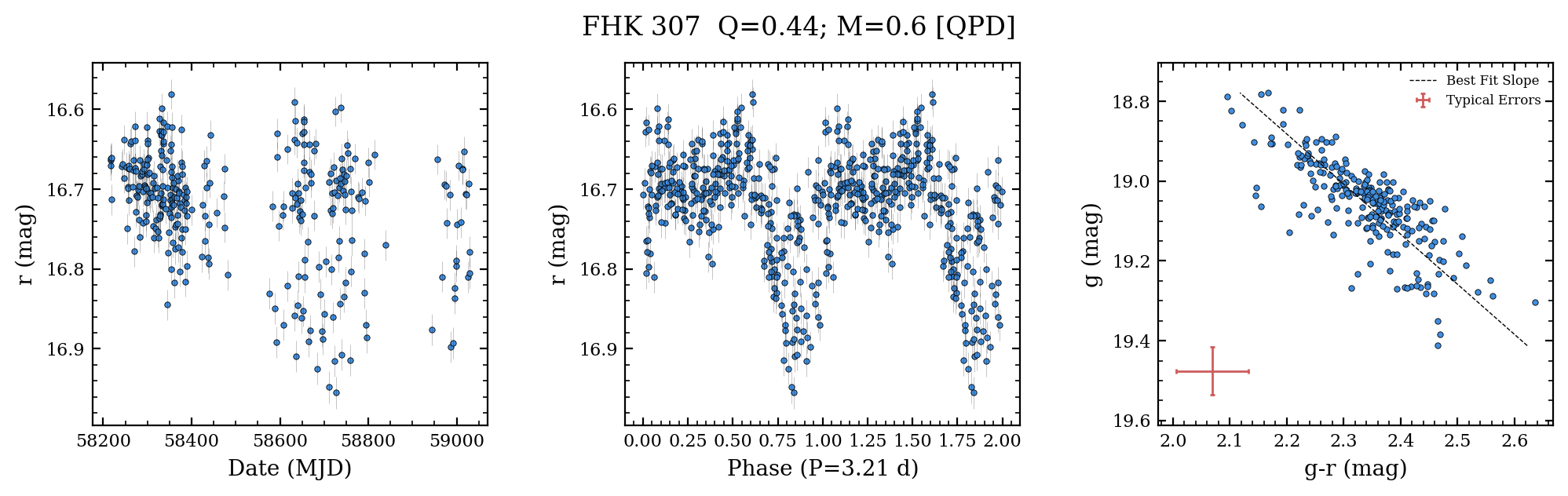}
\includegraphics[width=0.7\textwidth]{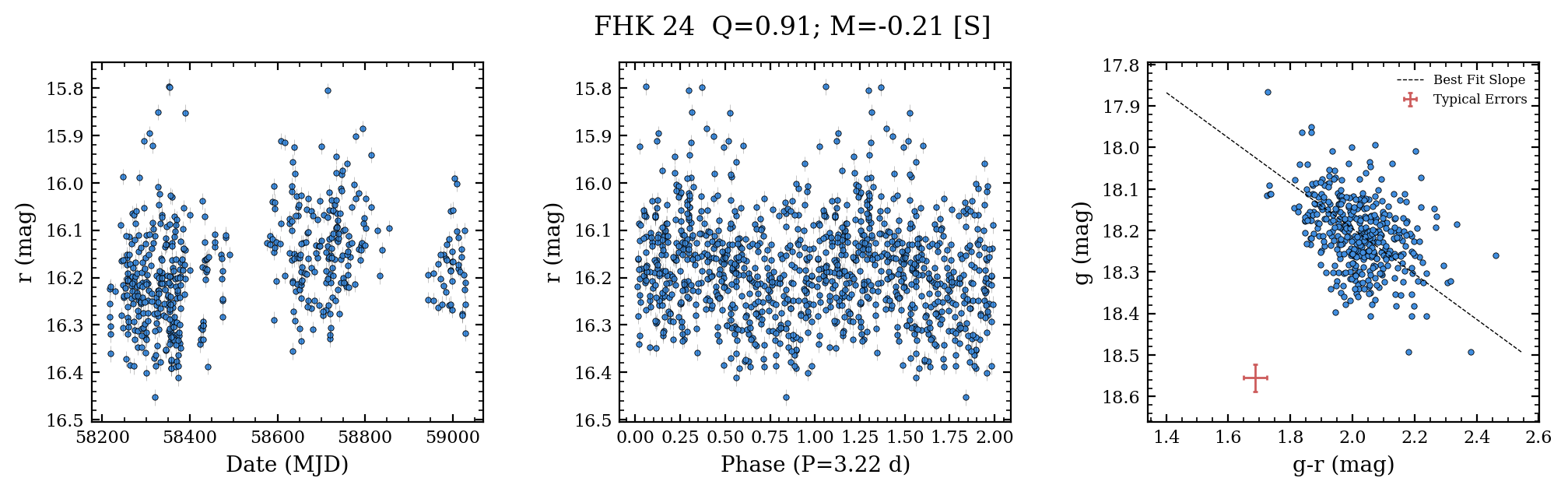}
\includegraphics[width=0.7\textwidth]{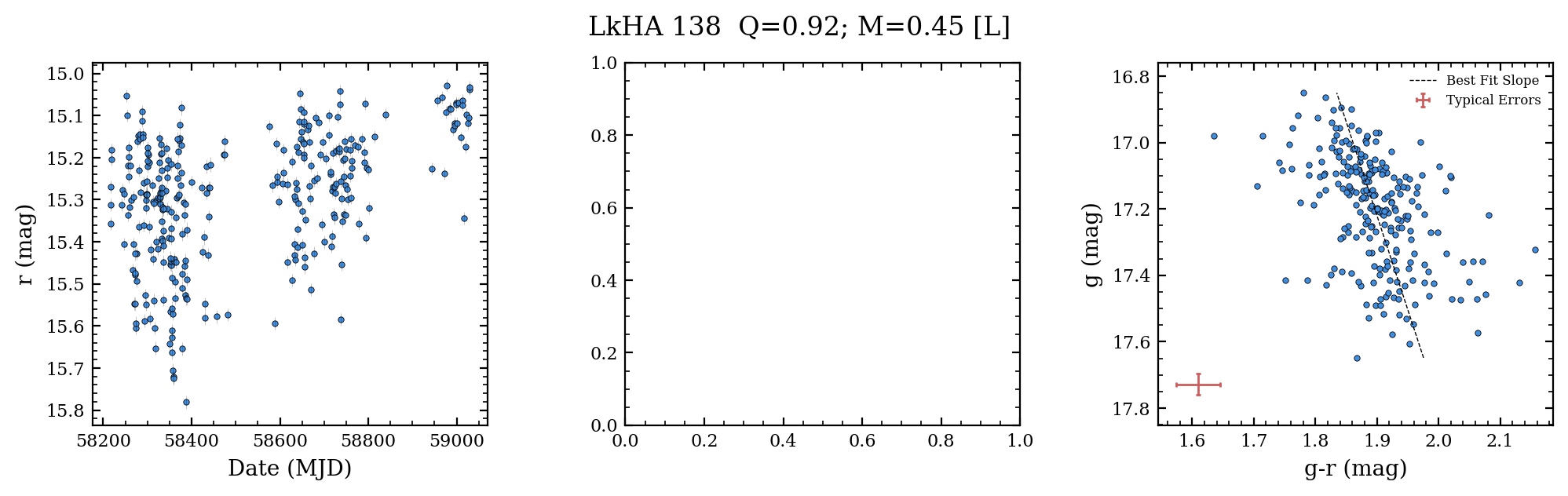}
\centering
\caption{Examples from the online figure set, illustrating objects from different parts of the $Q-M$ plane. One object from each of the periodic (P), quasi-periodic symmetric (QPS), quasi-periodic dipper (QPD),  stochastic (S), and long-timescale (L) groups is shown, complementing the aperiodic dipper (APD) and burster (B) categories shown in the main text Figures~\ref{slopesGallery} and ~\ref{slopesGallery2}.
In each panel, object name along with the Q and M values, the assigned light curve morphological class,
and the reported timescale are provided in the plot labelling.  
}
\label{examples}
\end{figure*}
In the main text, Figures~\ref{slopesGallery} and ~\ref{slopesGallery2}
presented exemplar objects from the ``dipper" and ``burster" categories, respectively. 
Figure~\ref{examples} now shows examples from the other morphological categories, including sources classified as ``periodic", ``quasi-periodic dipper", ``quasi-periodic symmetric", ``stochastic", and ``long timescale". 
An online supplemental figure set provides similar figures for every object in the variable star sample. 

For each of the 323 sources, we plot the light curve as well as the light curve folded to the time scale 
value reported in Table~\ref{tab:properties}, if one exists, and the $g$ vs $g-r$ color-magnitude diagram. 
If no period was detected by the Lomb-Scargle algorithm with power greater than the 99\% False Alarm Probability, 
then the folded light curve subplot is left blank. 
%Objects that are not classified as periodic variables with reported periods have periods of uncertain believability. 
If the slope angle error is less than ten degrees, then the best fit CMD slope also appears in the CMD subplots.

%%%%%%%%%%%%%%%%%%%%%%%%%%%%%%%%%%%%%%%%%%%%%%%%%%%%%%%%%%%%%%%%%%%%%%%%%%%%%%%%%%%%%%%%%%%%%%%%%%%%%%%%%%%%%%%%%%%%%%%%
\newpage

%\section{Anomalously Slow Rotators}  \label{slowRotators}
\section{A Lesson Regarding True Periods Near the Observing Cadence}  \label{slowRotators}

Two high-confidence NAP members are classified as strictly periodic (P) sources, but have periodograms with strong
peaks in an unusual pattern.  In each case, of the four peaks, several are related as beat aliases, 
but it is not clear which is the real period, which two are the beats, and which one is the odd extra period.  
Two of the peaks are just above and just below 1 day.
In addition, there are two more peaks around 40 days,
corresponding to unexpectedly slow stellar rotation rates; these long sinusoids are visible -- by eye -- in the light curves.  

\begin{figure}[htp]
    \centering
    \includegraphics[width=0.48\textwidth]{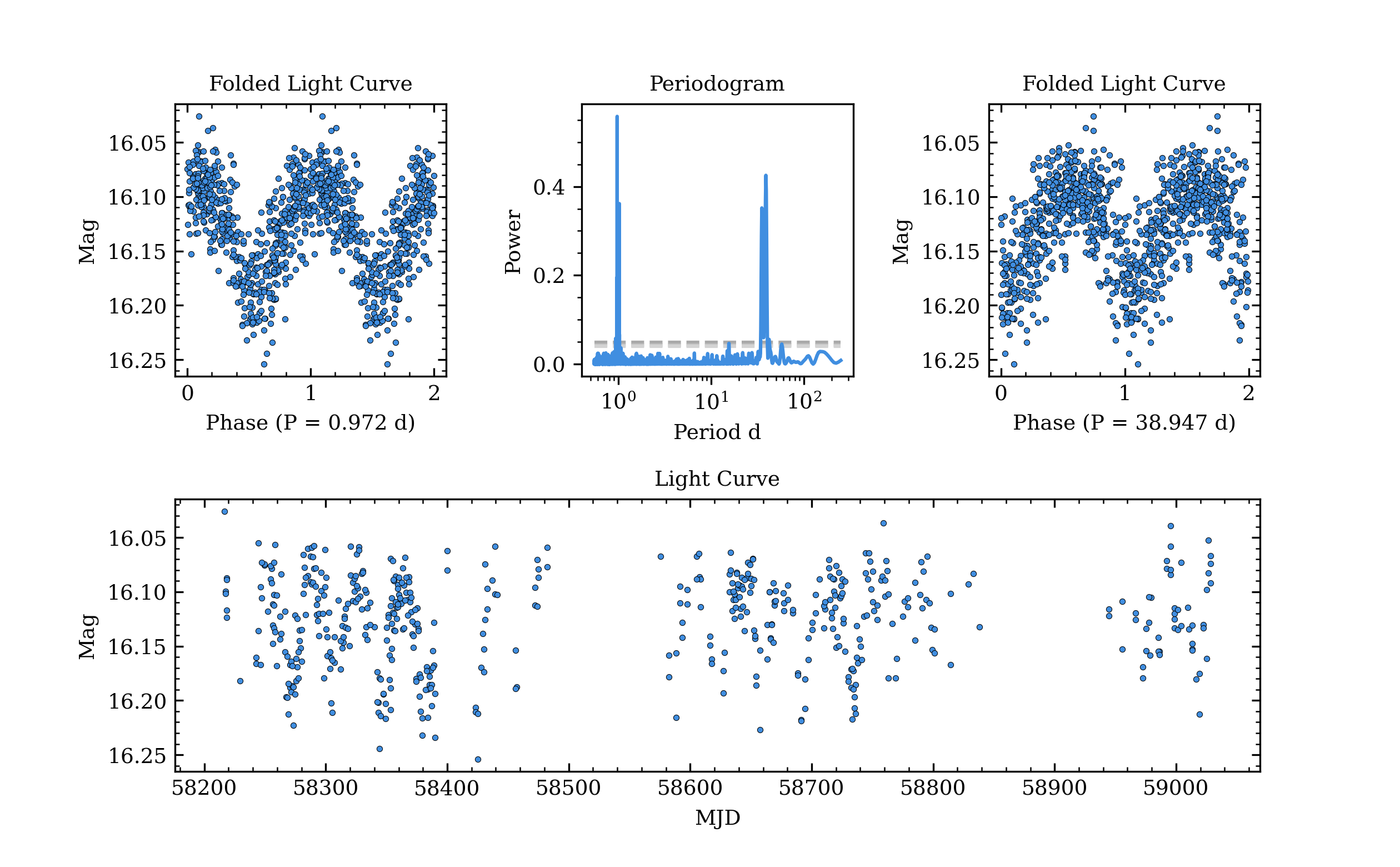}
    %\vskip 0.5truein
    \includegraphics[width=0.48\textwidth]{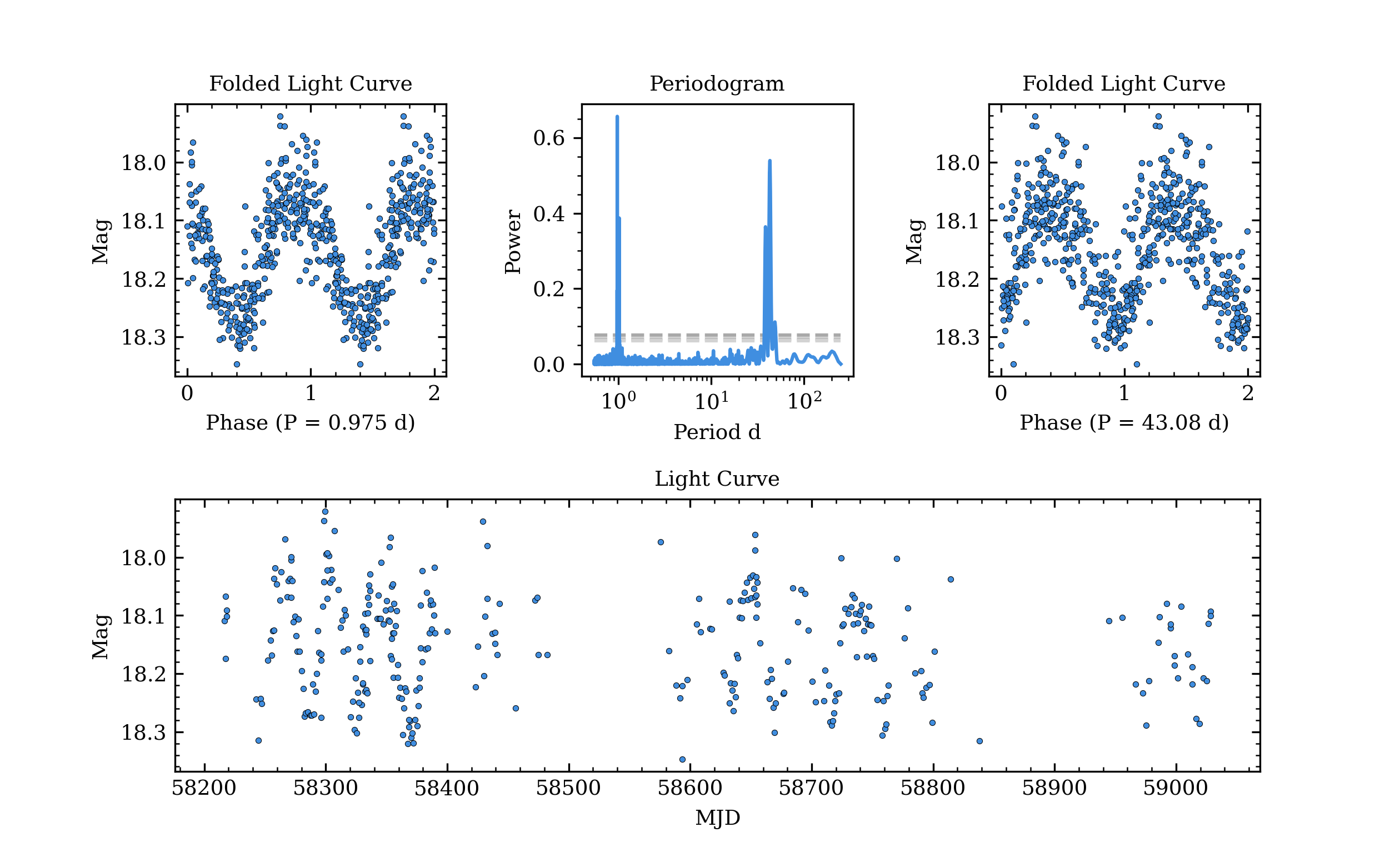}
    \caption{Light curves, periodograms, and phased light curves at the two strongest peaks for FHK 176 and FHK 286.
    Each of these sources has a complex periodogram structure that includes a double-peak around $\sim 1$ day and another double-peak around $\sim40$ day.  The latter periodic signal can be seen directly in the unphased light curves. 
    However, in both cases, the short period just below one day appears to be the real period. 
    See text for how the signal can be deciphered.}
    \label{fig:anomalous}
\end{figure}

The two sources are FHK 176 and FHK 286, with 
period amplitudes of $\sim0.1$ and $\sim0.2$ mag respectively, suggestive of significantly spotted photospheres.
Their time series and phased light curves  are illustrated in Figure~\ref{fig:anomalous}.  
Based on the information in \cite{Fang2020}, one star has spectral type K8.5 and the other M0; both have \ion{Li}{1} absorption 
and H$\alpha$ emission, plus securely pre-main sequence locations in the H-R diagram.  Neither star has identifiable infrared excess.

If real, the slow rotation rates of these two stars are far out on the tail of the period distribution (Figure~\ref{periodsdist}) for this cluster-- a full factor of 10 above the median rotation rate -- 
and anomalously slow for members of a star-forming region.
Indeed, it is unusual for late K-type stars of any age to rotate this slowly, although some such slow rotators are known among the field star population and among $>5$ Gyr old cluster members \citep{Curtis2020}.

On interpretation is that the multi-periodicity could indicate that these two sources are each binary systems,
in which both components of the binary are individually detected as spotted rotators.   In this scenario, the two
short periods might be supposed as real, with the two long periods beat aliases.  However, the relative periodogram
peak heights do not seem consistent with this.   
Furthermore, a recent Keck/HIRES spectrum shows no sign of a spectroscopic binary in cross correlation analysis.
If, instead, one of the short and one of the longer periods is real,
then we would need to explain how angular momentum was removed so quickly from just one component of a binary.
We ultimately reject the short$+$long period binary scenario on the grounds that subtracting just one of these signals
from the light curve effectively removes both peaks in the periodogram. 

In terms of beat periods where $1/P_{beat} = 1/P_{true} \pm 1$, the taller peak around 1 day has a beat at the shorter peak around 40 days.  Similarly, the taller peak around 40 days has a beat at the shorter peak around 1 day.  Although we can explain one of the longer periods as a beat, we can not explain both long periods in that manner through simple approximations.  

The situation resolves itself when we consider that the strongest peak for each of the two stars is just under one day,
and quite close to the data sampling interval.  
Simulating the short period at the actual ZTF cadence does produce both short-period and both long-period peaks in the observed ratio of their powers.
Specifically, injecting a pure sinusoid with period 
0.972 days (FHK 176) and 0.975 days (FHK 286) using the actual MJD values from the ZTF data stream, 
results in the long period peaks that can be seen by eye in Figure~\ref{fig:anomalous}.  
The long period aliases of the true short periods weaken substantially only when the sampling times are randomized
beyond the current staggered ZTF sampling in this particular field, with a standard deviation exceeding 4 hours.

As a check on this conclusion, we consulted the ZTF high cadence data \citep{Kupfer2021} available for this field
on MJD = 58448, 58451, and 58455 days.  Although stated earlier as having generally poor photometry, one of these nights
is better than the other two, and for FHK 176 the data support the tallest periodogram peak as the true period. 
Systematic changes in the star's brightness 
can be seen during the single night, suggesting that the short $\sim1$ day period is the correct one.  
For FHK 286, the situation is more ambiguous but we believe the same conclusion applies.

%Their ages are too young and their periods are too long for spin-orbit locking effects to be a viable explanation, however.  
%Instead, some other star-star or earlier stage star-disk -- star-disk interaction may have drained away angular momentum.
%One possible explanation is if a companion in a retrograde orbit spins the star down quickly
%\footnote{We thank Luke Bouma for suggesting this possible interpretation.}.
%As shown in models by \cite{Anderson2021} a retrograde binary plus tides can produce slow rotators (see their Figure 4), with a binary star even more advantageous than the hot Jupiter case for which the model was developed.

% see herbst2002 page 518 discussion and figure 6.

%%%%%%%%%%%%%%%%%%%%%%%%%%%%%%%%%%%%%%%%%%%%%%%%%%%%%%%%%%%%%%%%%%%%%%%%%%%%%%%%%%%%%%%%%%%%%%%%%%%%%%%%%%%%%%%%%%%%%%%%
\newpage

\section{The Influence of Cadence and Photometric Precision on Q}
The Q and M metrics were developed to quantify stellar variability in high-quality, regularly spaced, photometric time series data taken from space-based platforms.  In this work, we have applied them to lower-precision light curves sampled at irregular intervals,
taken from the ground.  We assess in this appendix the transferability of these metrics from space-based to ground-based data. To do so, we explore how $Q$ behaves on a $K2$ data set -- specifically the $\sim 100$ young stars in the Taurus region with light curves reported in \cite{Cody2022}, and versions of these data that have been degraded to the cadence and precision typical of our ground-based ZTF data set. 

For each light curve in the Taurus sample observed by $K2$, we began with a baseline $Q$ value that was calculated as in \S5 for the ZTF data.  Then, we down-sampled the observational cadence of the Kepler/K2 data set to match the actual time series intervals of the ZTF data set, and recalculated the $Q$ metric.  
We also recalculated the $Q$ metric based on down-sampled and error-adjusted Kepler light curves.
For this, we adopted error bars appropriate to the ground-based data, sampling from the empirical error distribution in Figure~\ref{fig:magmagerr}.  In practice we assumed two error values, 0.015 and 0.036 mag, which represent the 10th and 90th percentile errors for the ZTF time series data analyzed in this paper. 

%The infrequent observations of ZTF stars’ magnitude could affect the Q significantly. The Lomb Scargle algorithm could detect a period that isn’t there because of how few and spread out the observations are. Additionally, there may be bias in the selection of observations, as they weren’t taken at random, affecting the measurements of the variance of the light curve and residuals. To quantify the effect of infrequent non-randomized selection, we decided to get the Q for 90 Kepler objects and calculate the Q on the 90 Kepler objects’ down-sampled to match the ZTF cadence. The Kepler objects’ observations are taken every 30 minutes, close enough together to get an almost exact light curve. Therefore, the Q metric on the Kepler object will be almost exact, and the frequent observations allowed us to almost perfectly match the ZTF’s cadence of observations. 

\begin{figure*}[htpb]
\includegraphics[width=0.5\textwidth]{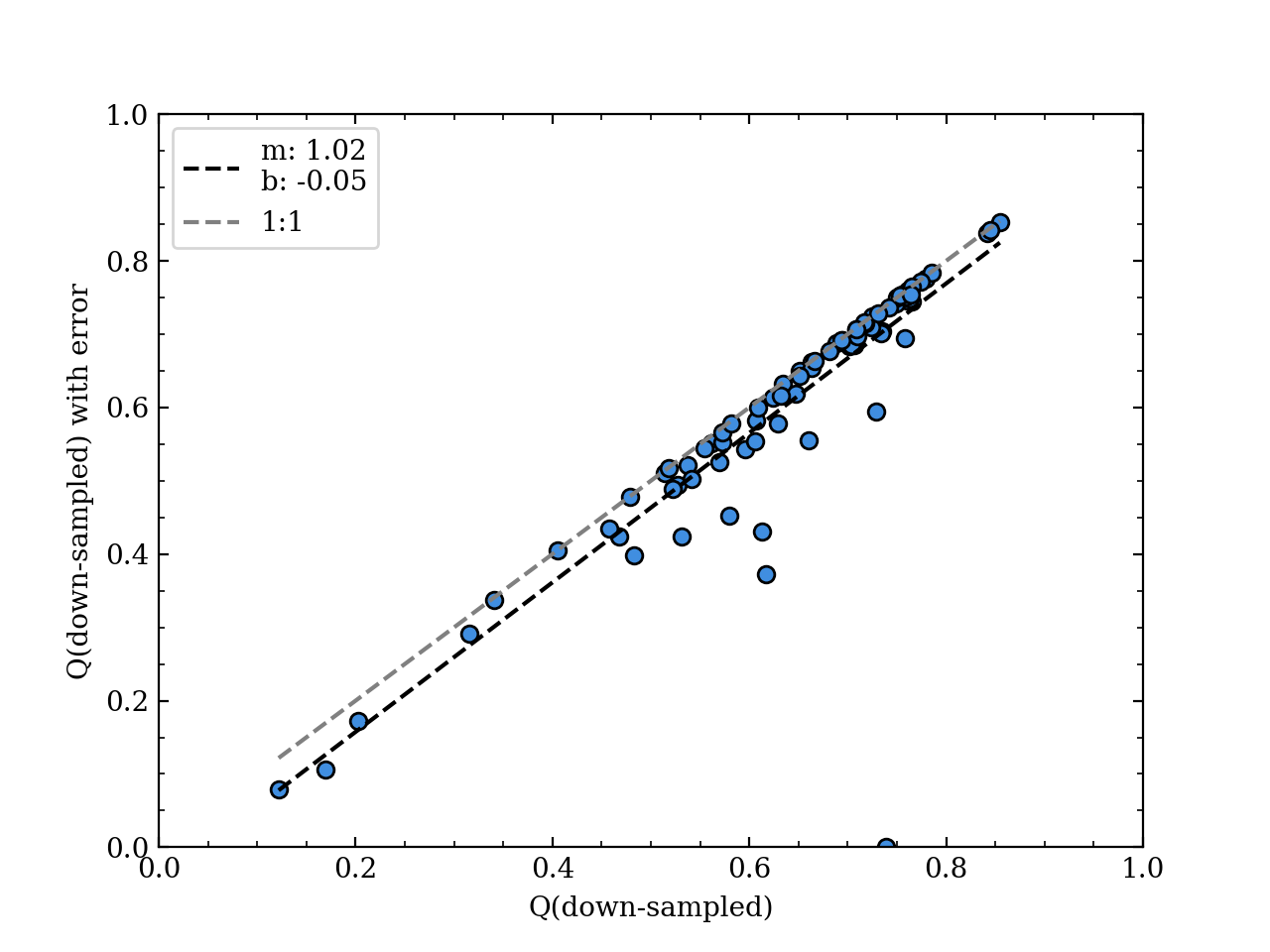}
\centering
\caption{Effect of light curve precision on the $Q$ metric.   Abscissa shows $Q$ values calculated in the absence of error, while ordinate shows $Q$ values for the same data but including realistic ground-based errors appropriate to our ZTF observations (0.036 mag).
Lines indicate the relationships shown in the legend.
Photometric error reduces $Q$ by an amount that likely depends on light curve amplitude relative to the photometric error, and light curve morphology. While the shift is systematic, many stars are unaffected and the slope remains close to unity.}
\label{test_precision}
\end{figure*}

\begin{figure*}[htpb]
\includegraphics[width=0.49\textwidth]{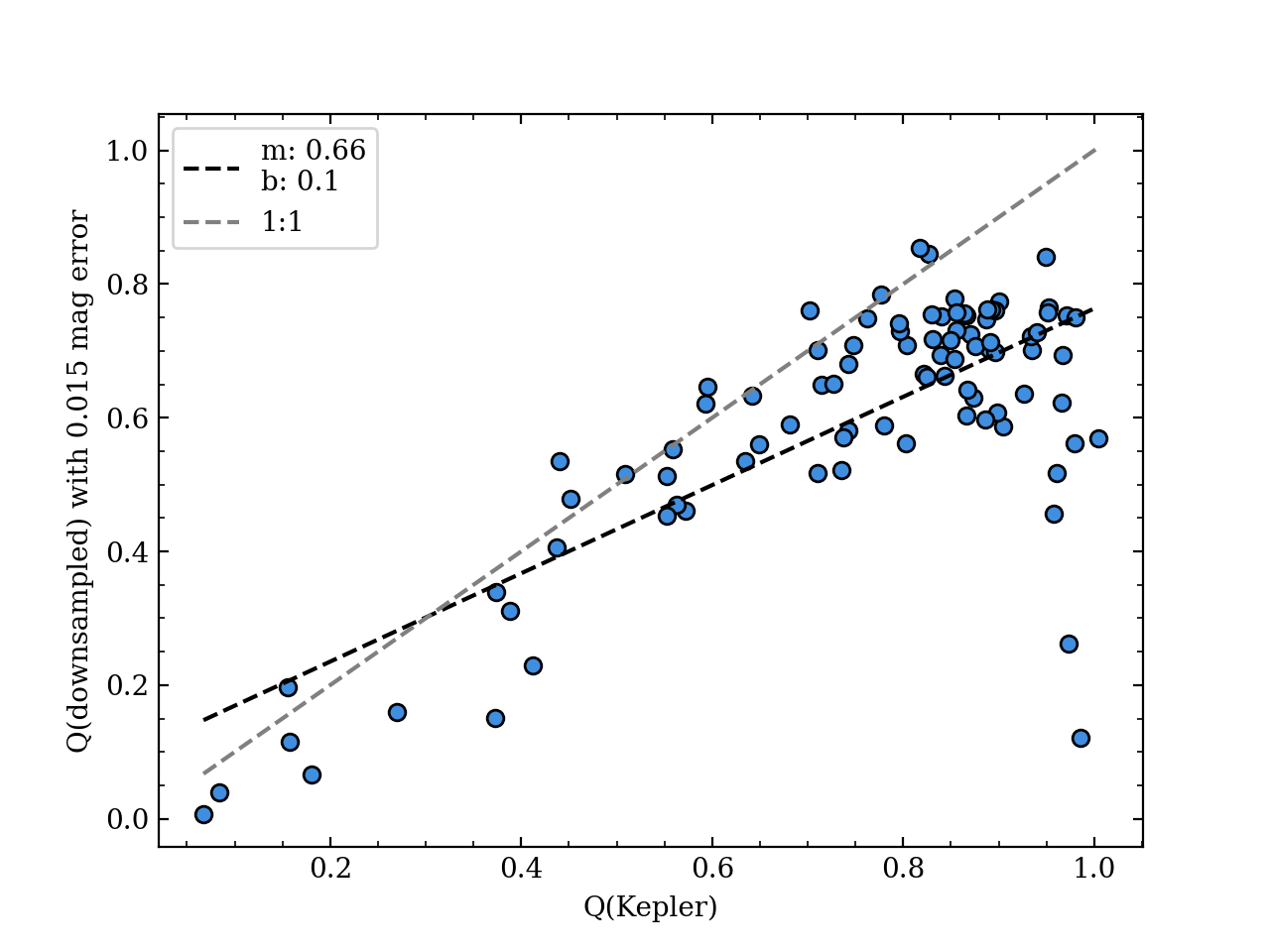}
\includegraphics[width=0.49\textwidth]{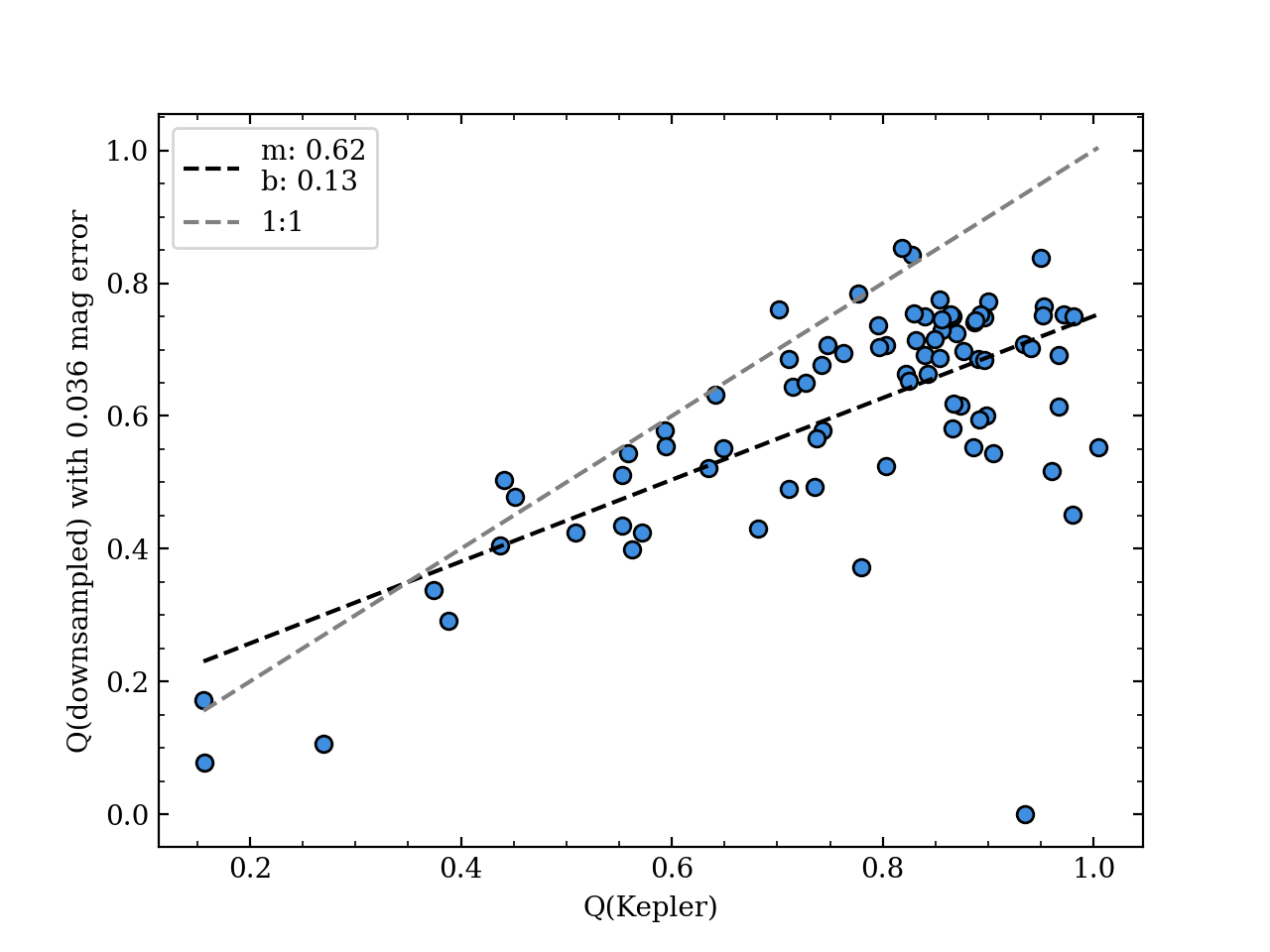}
\centering
\caption{Effect of light curve cadence on the $Q$ metric, illustrated for two different values of the light curve precision.
Abscissa shows $Q$ values calculated at the native cadence of $K2$, while ordinate shows $Q$ values for the same K2 data but down-sampled to the true cadence of our ZTF observations over the same limited time interval, with the indicated errors included. Lines indicate the relationships shown in the legend. 
Cadence has a large effect on Q, with more scatter towards higher $Q$.   The effects of error are apparent, with overestimated errors leading to a drop in $Q$. The values go negative for some stars, particularly in the higher error (right) panel, causing them to fall outside the bounds of the plot.}
\label{test_cadence}
\end{figure*}

Considering first the influence of photometric uncertainty on the derived $Q$,  Figure~\ref{test_precision} illustrates for the down-sampled data, that the introduction of error lowers the $Q$ values. This is the expected behavior since the photometric error is subtracted from both the numerator and the denominator in Eq.~\ref{eq:defq}. However, there are a few cases in which the variance of the residuals is so small, that the numerator goes negative. Similarly, the actual variance of the light curve can be smaller than the mean photometric error, causing $Q$ to be greater than 1.  
This happens because, in the $Q$ statistic, we are adding error only to the $\sigma^2_{phot}$ term and are not inflating the actual $\sigma_m$ and $\sigma_{resid}$ terms in Equation 2.  In these simulated cases, the astrophysical variability is smaller than the simulated error, leading to the unphysical values of $Q$. 

In the simulated ground-based data, including both the precision and the cadence adjustments from space to ground,
the range of $Q$ values extends from -3.18 to 2.20,  compared to 0.068 - 1.01 for the original high-cadence light curves.
We have excluded these unphysical values in the linear fit shown in Figure~\ref{test_precision}.
For the analysis below, we also consider only $Q$ values between the nominal 0-1 range.

Now considering the influence of cadence on the derived $Q$,
in Figure~\ref{test_cadence} we compare $Q$ from the mock ground-based data set (down-sampled, with added errors) to $Q$ from the original, fully sampled and essentially error-free Kepler data set.  While there is a clear positive association remaining, the relationship is not one-to-one, as above when only error was considered. Instead, the relationship flattens out as $Q$ increases. This can be attributed mostly to the Lomb Scargle algorithm detecting periods that are not actually present, which then affects the calculation of the light curve residual in the Q formalism (Eq.~\ref{eq:defq}). 

We note that the downsampling of the Kepler data stream takes into account only a fraction of the number of data points that were available for calculating $Q$ in the full ZTF time series. A major consideration in this experiment is that the Kepler objects were measured over only about 80 days, while the ZTF objects were observed over 300 days (including data through ZTF DR4). Thus, the actual ZTF periodograms are based on more data than the down-sampled Kepler periodogram analysis. Therefore, the ZTF periodograms and hence the residual light curves are probably a more accurate representation of the period-subtracted residual plot for objects with long timescales. 
We found that the $Q$ values are fairly accurate when they are low, but became increasingly more inaccurate as the actual Kepler Q increased. This finding should translate over to the actual ZTF objects, i.e. lower $Q$ values are more accurate than higher $Q$ values. 
Additionally, all of the down-sampled Q values were slightly underestimated, which again is probably true of the ZTF objects as well. 

Overall, this experiment reveals that there are constraints on data quality for assessing $Q$ in various photometric data sets.
Specifically, while photometric uncertainty is important, lower cadence has a more pronounced affect on $Q$, substantially reducing it for certain categories of variability.

\end{appendix}

\end{document}